\documentclass[conference]{IEEEtran}
\pdfoutput=1

\newcommand\blfootnote[1]{%
  \begingroup
  \renewcommand\thefootnote{}\footnote{#1}%
  \addtocounter{footnote}{-1}%
  \endgroup
}

\usepackage{amsmath,amssymb,amsfonts}
\usepackage{graphicx}
\usepackage{textcomp}
\usepackage{xcolor}
\usepackage{amsmath}
\usepackage{subfig}
\usepackage{enumitem}
\usepackage{algorithm}
\usepackage{algpseudocode}
\usepackage{multirow}
\usepackage[flushleft]{threeparttable}
\usepackage{siunitx}
\usepackage{authblk}

\usepackage{tabularx}
\newcounter{protocol}
\newenvironment{protocol}[1]
  {\par\addvspace{\topsep}
   \noindent
   \tabularx{\linewidth}{@{} X @{}}
    \\\hrule
    \refstepcounter{protocol}\textbf{Protocol \theprotocol} #1 \\
    \hrule
    }
  { 
   \hrule
   \endtabularx
   \par\addvspace{\topsep}}
\newcommand{\sbline}{\\[.5\normalbaselineskip]}
\def\BibTeX{{\rm B\kern-.05em{\sc i\kern-.025em b}\kern-.08em
    T\kern-.1667em\lower.7ex\hbox{E}\kern-.125emX}}
    
\begin{document}

\title{DeepAuditor: Distributed Online Intrusion Detection System for IoT devices via Power Side-channel Auditing }


%
%
\author[1]{Woosub Jung}
\author[2]{Yizhou Feng}
\author[2]{Sabbir Ahmed Khan}
\author[2]{Chunsheng Xin}
\author[2]{Danella Zhao}
\author[1]{Gang Zhou}
\affil[1]{Department of Computer Science, William \& Mary}
\affil[2]{Department of Computer Science, Old Dominion University}

%


\maketitle

\begin{abstract}
As the number of IoT devices has increased rapidly, IoT botnets have exploited the vulnerabilities of IoT devices. However, it is still challenging to detect the initial intrusion on IoT devices prior to massive attacks. Recent studies have utilized power side-channel information to identify this intrusion behavior on IoT devices but still lack accurate models in real-time for ubiquitous botnet detection.

We propose the first online intrusion detection system called DeepAuditor for multiple IoT devices via power auditing. To develop the real-time system, we propose a lightweight power auditing device called Power Auditor. We also design a distributed CNN classifier for online inference in a laboratory setting. In order to protect data leakage and reduce networking redundancy, we then propose a privacy-preserved inference protocol via Packed Homomorphic Encryption and a sliding window protocol in our system. The classification accuracy and processing time are measured, and the proposed classifier outperforms a baseline classifier, especially against unseen patterns. We also demonstrate that the distributed CNN design is secure against any distributed components. Overall, the measurements are shown to the feasibility of our real-time distributed system for intrusion detection on IoT devices. 
\end{abstract}

\blfootnote{Preprint. Accepted by IPSN'22, May 3-6, 2022, Milan, Italy}
\section{Introduction}\label{section:1}
Internet of Things (IoT) devices have become the new cybercrime intermediaries between attackers and users to process cyberattacks. For example, in October 2016, a massive distributed denial-of-service (DDoS) attack incapacitated the Domain Name System provider Dyn. This made several Internet platforms and services, such as Amazon, Netflix, PayPal, and Twitter, temporarily unreachable to numerous users in Europe and North America. This IoT botnet attack was called Mirai and exceeded 600 Gbps in volume at its peak. This overwhelming amount of the traffic was sourced from 65,000 injected IoT devices, including routers, security cameras, and digital video recorders \cite{Anonymous:dcm0bUHc}. These IoT devices were known at the time to have weak security protection and to be vulnerable to attacks. As reported by Symantec \cite{web:symantec}, thousands of outdated routers were targeted by the worms exploiting their vulnerability. Since then, many variants of Mirai have emerged to target the weaknesses of IoT devices. Besides serving as the intermediaries of DDoS attacks, IoT devices were also found to serve as attack proxies for multiple cybercrimes, such as clickjacking and spearphishing \cite{iot:phishing}\cite{Panwar:2019uo}.

Even though the Mirai attack happened five years before the time of writing, IoT devices are still vulnerable to IoT botnets. Mirai and its variants used a simple brute-force attack \cite{brute} when first transforming IoT devices into their bots, which is still cybercriminals’ preferred option for executing attacks \cite{survvv}. Despite the clear intrusion procedures of botnet attacks on IoT devices, it is not easy to determine whether an intrusion has occurred. The main concern is that the network traffic generated on endpoint devices is not noticeable as malicious behavior in the initial stages of the attack. In this situation, deploying network-based botnet detection systems into different IoT devices/vendors is heavyweight and intrusive to users. For example, 84 different IoT devices/vendors were found to engage in the Mirai bots. Moreover, they were related to more than 300 different communication protocols and platforms \cite{Anonymous:dcm0bUHc}. Thus, network-based botnet detection requires a great deal of modification in programming languages or operating systems on diverse IoT devices. One approach to tackling this issue involves the use of power side-channel information for the detection of stealthy attacks. Using a power side-channel is durable and universal since power traces are hard to compromise and can capture accumulated tasks on heterogeneous devices, such as different hardware, vendors, operating systems, etc. Meanwhile, it is nearly impossible for adversaries to mimic normal power draw behavior while attacks. Thus, some pioneering work in this area utilized power side-channel data to detect malignant behavior on mobile devices in the early 2010s \cite{Kim:2008cl}\cite{Yang2017OnIB}. However, these studies are outdated and need to be validated in IoT environments.

Several recent studies have utilized power auditing — the analysis of power consumption — to explore IoT security and defend against malicious attacks \cite{Myridakis:2021tg}\cite{Li:ej}. However, those proposed systems mainly focused on detecting massive DDoS attacks on IoT devices. It is still therefore challenging to distinguish the subtle initial stages of an IoT botnet intrusion. Jung et al. \cite{Jung-01} introduced IoT botnet detection via power modeling. In this research, the authors designed a Convolutional Neural Network (CNN) model using one-dimensional power side-channel data to classify malicious intrusion. While the CNN classifier showed promise in detecting subtle differences in power traces, the fundamental problem with this study is that it was conducted offline with a bulky and expensive power monitor. Thus, it is not practical for ubiquitous botnet detection on IoT devices. To tackle these challenges, we designed a lightweight power auditing device and a distributed online CNN classification system for resource-constrained IoT devices. Our proposed end-to-end system \textit{DeepAuditor} was developed in a distributed setting and can simultaneously detect the initial botnet intrusion on multiple IoT devices via power auditing.

With this aim, our research questions in this paper are as follows:
\begin{itemize}
\item How can we audit power side-channel information of IoT devices in real-time for ubiquitous botnet detection?
\item How can we conduct online inference for multiple IoT devices in a distributed setting? 
\item How can the distributed classifier prevent data leakage and networking redundancy?
\end{itemize}

To answer the first question, we designed a small form-factor device called Power Auditor that is capable of measuring the power traces of a connected IoT device for behavior classification. Unlike off-the-shelf power monitors, our Power Auditor is lightweight and portable. Thus, we envision devices such as Power Auditor as an important component of future smart plugs. Nowadays, smart plugs are often used in IoT environments to control electronics remotely for the sake of convenience, e.g., Amazon Smart Plug \cite{web:smpg}. As IoT devices are getting popular, the smart plug market also grows quickly; its compounding annual growth rate is approximately 42\% over the forecast period between 2021 and 2026 \cite{smartplug_survey}. According to our prototype performance, the proposed algorithms are lightweight and therefore can easily be integrated into future smart plugs for ubiquitous botnet detection.

To address the other research questions, we developed distributed CNN classifier components: \textit{Data Inferencer} in a user site and \textit{Computing Cloud} in a cloud site. The proposed CNN design outperforms the baseline classifier; we increased the classification accuracy by up to 17\% in leave-one-out tests. In our leave-one-device-out tests, a classifier is trained with a certain device-type dataset and tested with the remaining device-type data to show its robustness against a new device. The leave-one-botnet-out tests are also crucial concerning practical deployments because the classifier has to be robust against unseen attack patterns but not overfitted. We then designed distributed protocols between our CNN components: a privacy-preserved CNN protocol and a sliding window protocol. Note that side-channel information can also reveal users' private data to the cloud site \cite{Maiti:2019vc}. To remedy this concern, we designed the privacy-preserved CNN protocol in our distributed system. This protocol also addresses the problem of attackers extracting classifier model parameters from distributed environments \cite{Zhang} \cite{Shokri}. The sliding window protocol was developed to reduce networking redundancy.


In summary, our contributions in this study are threefold.
\begin{itemize}
\item As a proof of concept, we designed a dedicated small form-factor device called Power Auditor to realize online botnet detection. The Power Auditor is lightweight and portable, and therefore can easily be integrated into future smart plugs. 

\item We are the first to develop a distributed online intrusion detection system for IoT devices via power auditing. We designed distributed CNN components and proposed distributed protocols between user and cloud sites in order to conduct real-time inference without data leakage.

\item 
We demonstrated the performance of our classification system in a laboratory setting: 1) The proposed CNN classifier detects malicious behavior with an accuracy of up to 98.9\%, which outperforms the baseline, 2) we theoretically analyzed the data protection of the privacy-preserved protocol, and 3) the distributed system supports about one hundred IoT devices simultaneously by using our laboratory server.
\end{itemize}

The rest of the paper is organized as follows: Section \ref{section:2} provides background and a threat model for our study. In Section \ref{section:3}, we introduce our botnet detection system DeepAuditor. In Section \ref{section:4}, we present the Power Auditor design of our system. Section \ref{section:5} introduces the distributed online CNN model for botnet detection. In Section \ref{section:implementation}, we describe online system implementation. Section \ref{section:7} demonstrates online performance evaluation of our system. Section \ref{section:8} presents our thoughts regarding limitations and future work. Section \ref{section:9} summarizes related work. Finally, we conclude this paper in Section \ref{section:10}.

\section{Background and Threat Model} \label{section:2}
This section provides background knowledge and introduces a threat model.

\begin{figure}[t]
\vspace{-5mm}
    \centering
      \subfloat[Device Rebooting ]{%
        \includegraphics[width=0.24\textwidth]{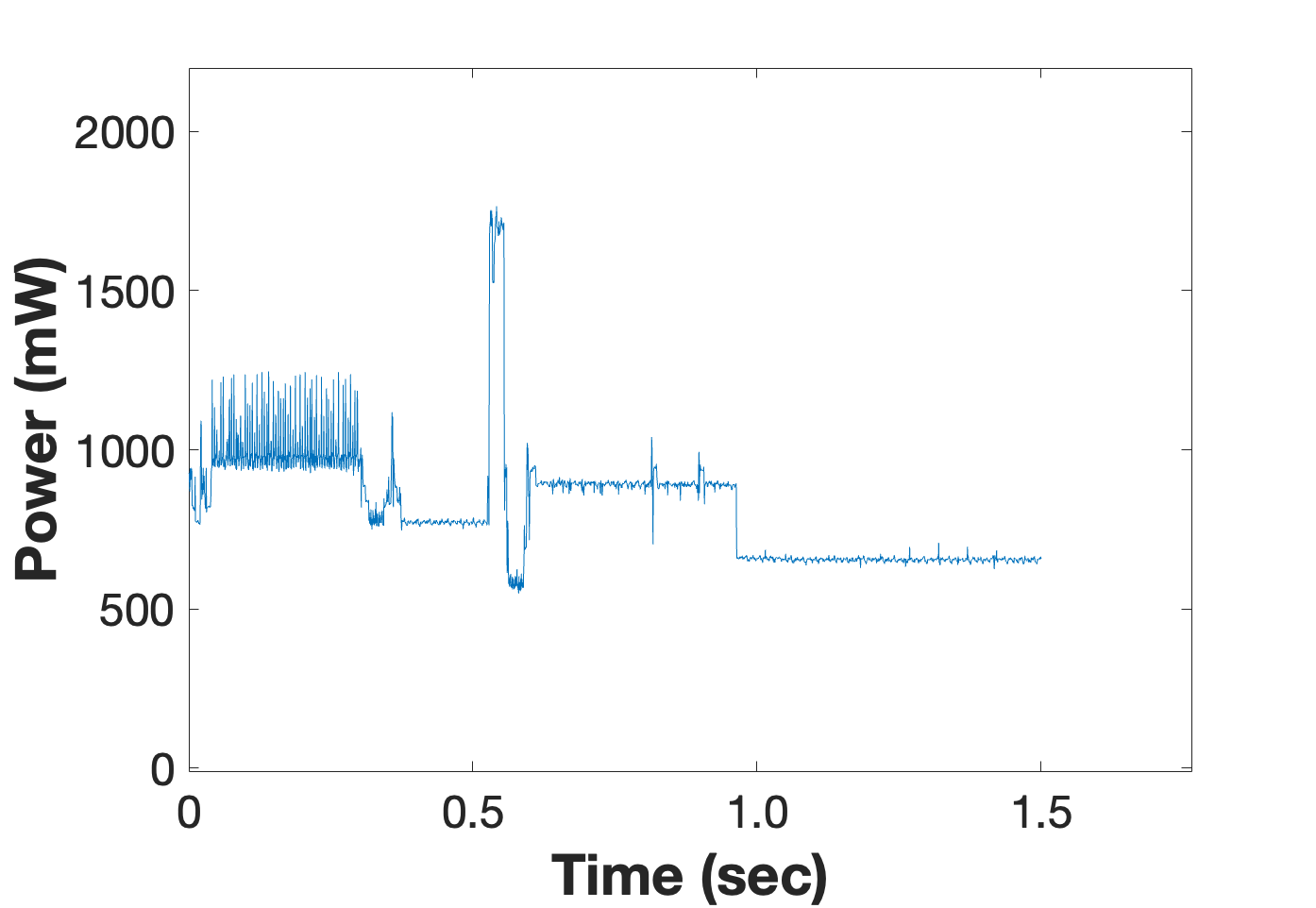}
        \label{fig:1a}}
      \subfloat[Botnet Intrusion]{%
        \includegraphics[width=0.24\textwidth]{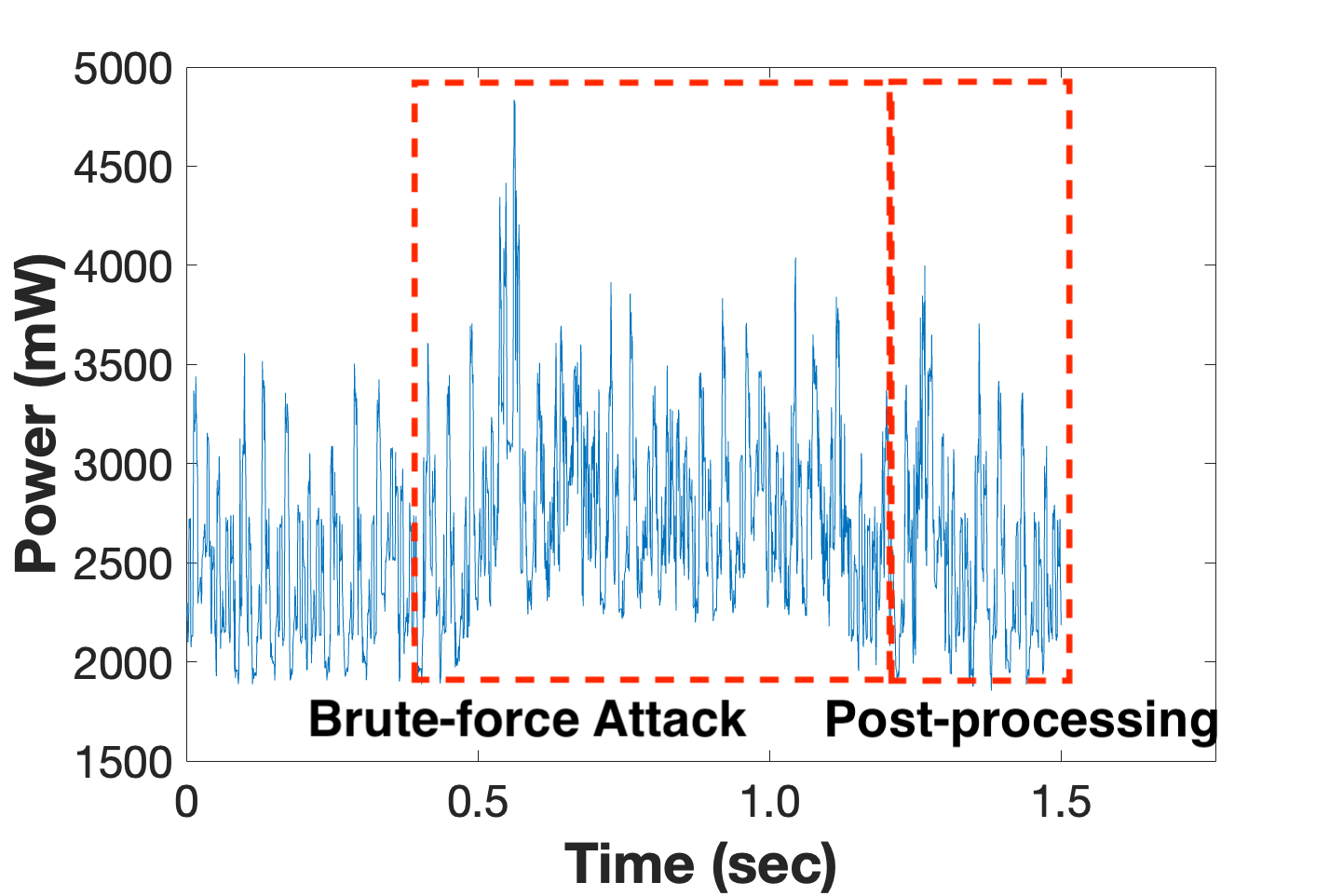}
        \label{fig:1b}}
    \caption{Power Traces collected by our Power Auditor } \label{fig:1}
    
    \vspace{-5mm}
\end{figure}

\begin{figure*}[t]
	\centering
    \includegraphics[width=0.96\textwidth]{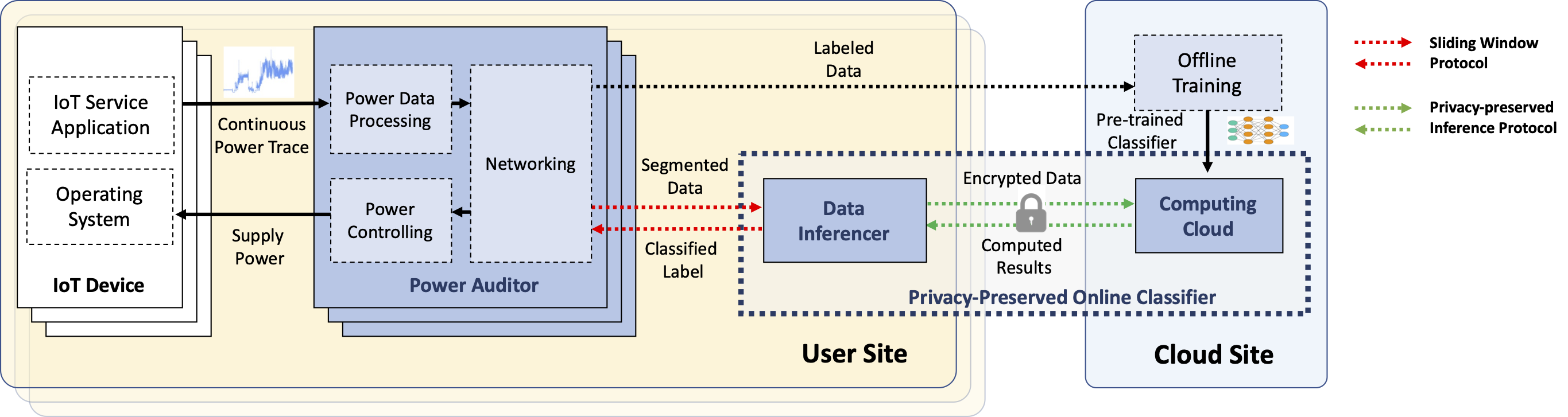}
	\caption{\textit{DeepAuditor} System Overview \textemdash\enspace The system consists of three subsystems with two distributed protocols.}
	\label{fig:2}
	\vspace{-3mm}
\end{figure*}

\subsection{Intrusion Detection via Power Modeling} \label{section:2.1}
Power traces can capture accumulated tasks to identify abnormal behavior. Several recent studies demonstrated that subtle activities on smartphones can be inferred from power consumption data \cite{Yang2017OnIB, qi:2018tn, Kim:2008cl}. Likewise, other studies on IoT security \cite{Jung-01}\cite{Li:ej} explored how IoT devices' behavior generates different patterns of power consumption data. For example, Figure \ref{fig:1} illustrates two examples of different activities collected by our Power Auditor. As shown in Figure \ref{fig:1a}, rebooting IoT devices generates power traces that are distinct from the power patterns that occur in Mirai botnet intrusion [Figure \ref{fig:1b}]. In Figure \ref{fig:1b}, an attacking bot enters the device using different username/password combinations and conducts post-processing. This generates two power traces: one from the brute-force attack and one from the post-processing. Furthermore, since the Mirai botnet family utilizes brute force attacks \cite{BF}, the power traces generated by these intrusions look nearly identical across dataset. Thus, it is feasible to identify malign behavior by analyzing power traces on IoT devices. 

In IoT botnet detection, it is crucial to detect this intrusion behavior in its early stages. Otherwise, botnets grow so quickly, so it will be too late to defend against DDoS attacks. We aim to identify whether the real-time power traces suggest malicious or benign behavior and classify them accordingly. Though RNN could also help in streaming data, it is more suitable in natural language applications. Instead, CNN is widely used in sensor streaming applications and classifies instances very fast in the inference stages. 
Therefore, we designed a CNN design to realize an online botnet intrusion detection system. More detailed design factors will be discussed in the following sections.

%

\subsection{Threat Model} \label{section:2.2}
\subsubsection{Client-side Vulnerabilities}
In DeepAuditor, we train power traces generated by existing IoT botnets. However, we also assume that an adversary is capable of conducting various patterns of botnet attacks. Thus, we consider two possible strategies that the adversary can use to attack our client-side. 1) Exploit vulnerabilities in the Power Auditor device. 2) Generate complicated post-processing jobs that create unseen power patterns. For example, downloading unexpected files or connecting to multiple servers can create more complicated data patterns.

To minimize the vulnerabilities of the Power Auditor, our proposed Power Auditor only monitors the power consumption of the physically connected IoT device within a local network; thus, the Power Auditor does not allow any inbound traffic from remote sources. This assumption is especially valid when smart plugs also do not allow users to access ssh/telnet services \cite{smartplug_survey}. Instead, these smart plugs are mostly managed by manufacturer apps. As we envision Power Auditor to be an integral part of future smart plugs, both Power Auditor and the smart plugs are secure against brute-force attacks. To address the post-processing job vulnerability, segmented data from different patterns were trained as botnet instances in our deep learning model. Thus, as long as power side-channel information is noticeable enough to label, our system is capable of detecting a diverse set of botnets beyond that are well-known. Our leave-one-out tests also demonstrate the classifier performance against unseen patterns in Section \ref{section:5.1}.

\subsubsection{Server-side Vulnerabilities}
In addition to exploiting the above vulnerabilities, adversaries could also target our cloud servers. Note that we implemented our classification model into two cloud-based edges. Thus, we assume any user-hold application and model-hold server in our system can become a semi-honest adversary. This means that they may try to steal information from received messages. For example, servers may infer IoT device behavior based on the power trace input, and users try to learn the server's model parameters based on the server output. We consider all parties non-colluding for their input data and output data. It is essential for our system to avoid user's privacy data disclosure that leads to poor credibility.

The emerging attacks presented in \cite{Zhang, Shokri} are also threats that we need to consider in our model. User-side can launch the model extraction attack \cite{Zhang} to extract the convolution layer and fully connected parameters based on the server received message. Server can process membership inferences attack \cite{Shokri} to compare the user input with the server's pre-trained dataset. 
In this research, our privacy-preserving mechanism masks the intermediate/final output for both user and server. However, the user still can learn the correct predicted result. Simultaneously, our privacy-preserving mechanism protects the server holds model parameters from the user, and user input is oblivious for server. 
We applied a flexible method to protect the output correctness and prove our system security by using a real-ideal paradigm \cite{Goldreich}, as introduced in detail in Section \ref{section:6.3}.

\section{\textit{Deep Auditor} System Overview} \label{section:3}

In this section, we introduce the distributed IoT botnet detection system DeepAuditor. As shown in Figure \ref{fig:2}, the proposed system consists of three subsystems: Power Auditor, Data Inferencer, and Computing Cloud. The ultimate goal of DeepAuditor is to identify subtle power differences in real-time between normal behavior and IoT botnet intrusions on multiple IoT devices. To build the system, Power Auditors are used to secure power-trace data of IoT devices. Data Inferencer and Computing Cloud are then used as online classifiers that can process intrusion detection. Altogether, the proposed distributed components accomplish the online intrusion detection via power side-channel auditing.

In the user site, the Power Auditor is connected to each IoT device to collect power consumption traces; there can be multiple Power Auditors for the corresponding IoT devices. We propose several functionalities of the Power Auditor for intrusion detection as follows. First, the Power Auditor is capable of auditing a device's power consumption footprint via the Power Data Processing module. During the online phase, the Networking module then transmits the collected data to the Data Inferencer for online classification. Finally, the Power Controlling module supplies power to the connected IoT device. Section \ref{section:4.1} introduces the universal design of the Power Auditor in detail.

Next, we developed a sliding window protocol between the Power Auditor and the Data Inferencer in the user site. This is because overlapping input instances of the CNN classifier can help to achieve better detection accuracy. However, if clients send those overlapping windows to the server sides, this will lead to networking redundancy. For example, if a Power Auditor sends 1.5 seconds of data every 0.5 seconds, this will bring 67\% networking redundancy. Instead, Power Auditors send segmented data to the Data Inferencer that concatenates the segmented packets for assembling input instances. Overall, in order to meet the real-time classification, our DeepAuditor needs to process each input instance less than the size of the sliding window. More details will be described in Section \ref{section:4.2}.

We then designed a one-dimensional CNN classifier for botnet intrusion detection, which we explain in Section \ref{section:5.1}. Our evaluation results demonstrate that our classifier outperforms the state-of-the-art classifier \cite{Jung-01} for power modeling. Based on the proposed 1-D CNN model, we implemented and deployed the pre-trained CNN classifier into a distributed lab setting.

Lastly, Section \ref{section:ppip} introduces the Privacy-Preserved CNN inference protocol for online prediction in a distributed setting. This protocol enables the CNN communications between user and cloud sites without leaking data. The Data Inferencer runs in the user site, receives power traces from Power Auditors, and sends encrypted data to the Computing Cloud to offload computation resources. In the cloud site, the Computing Cloud is in charge of CNN inference computations, i.e., nonlinear activation computations in the CNN. Then, the Computing Cloud sends the encrypted computation results to the Data Inferencer for online inference. Finally, the Data Inferencer classifies the final prediction of whether the given input power trace is benign or malicious.

\begin{figure}[t]
	\centering
    \includegraphics[width=0.31\textwidth]{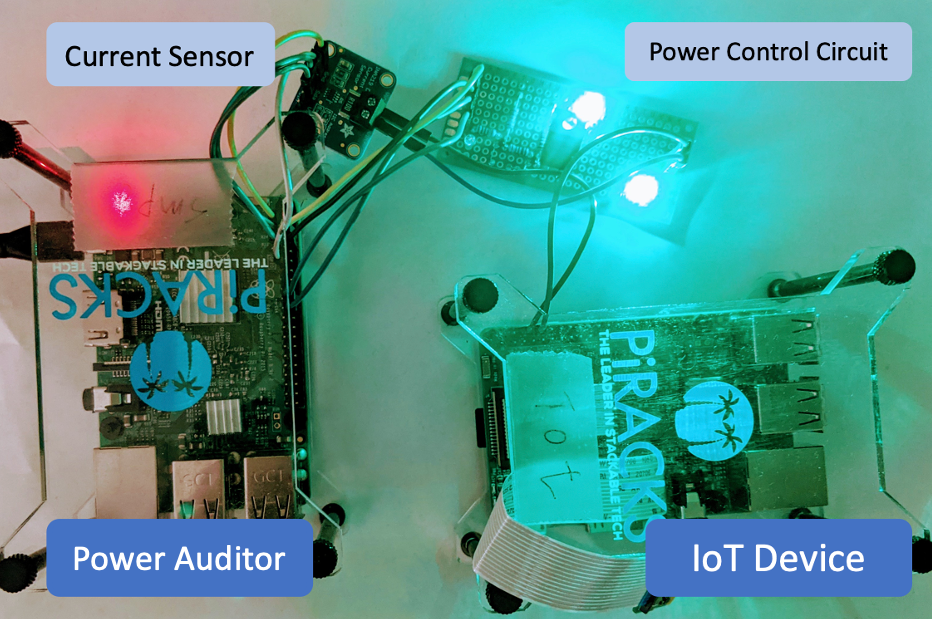}
	\caption{Power Auditor Prototype}
	\label{fig:auditor}
\vspace{-5mm}
\end{figure}

\section{Power Auditor Design} \label{section:4}

In section \ref{section:4.1}, we introduce the three universal software components of the Power Auditor: Power Controlling, Power Data Processing, and Networking modules. In section \ref{section:4.2}, we then describe the data transmission protocol between the Power Auditor and the Privacy-Preserved CNN Classifier.

\subsection{Component Design}\label{section:4.1}

In our Power Monitor, the Power Controlling module supplies power to a connected IoT device. Next, the Power Data Processing module reads power consumption traces of the attached IoT device. The Networking module communicates with the server-side to convey the sensing data for online classification. 
Figure \ref{fig:auditor} shows our prototype of a Power Auditor and a connected IoT device. All the source codes for Power Auditors will be available for download.

\begin{figure}[t]
	\centering
		\includegraphics[width=0.35\textwidth]{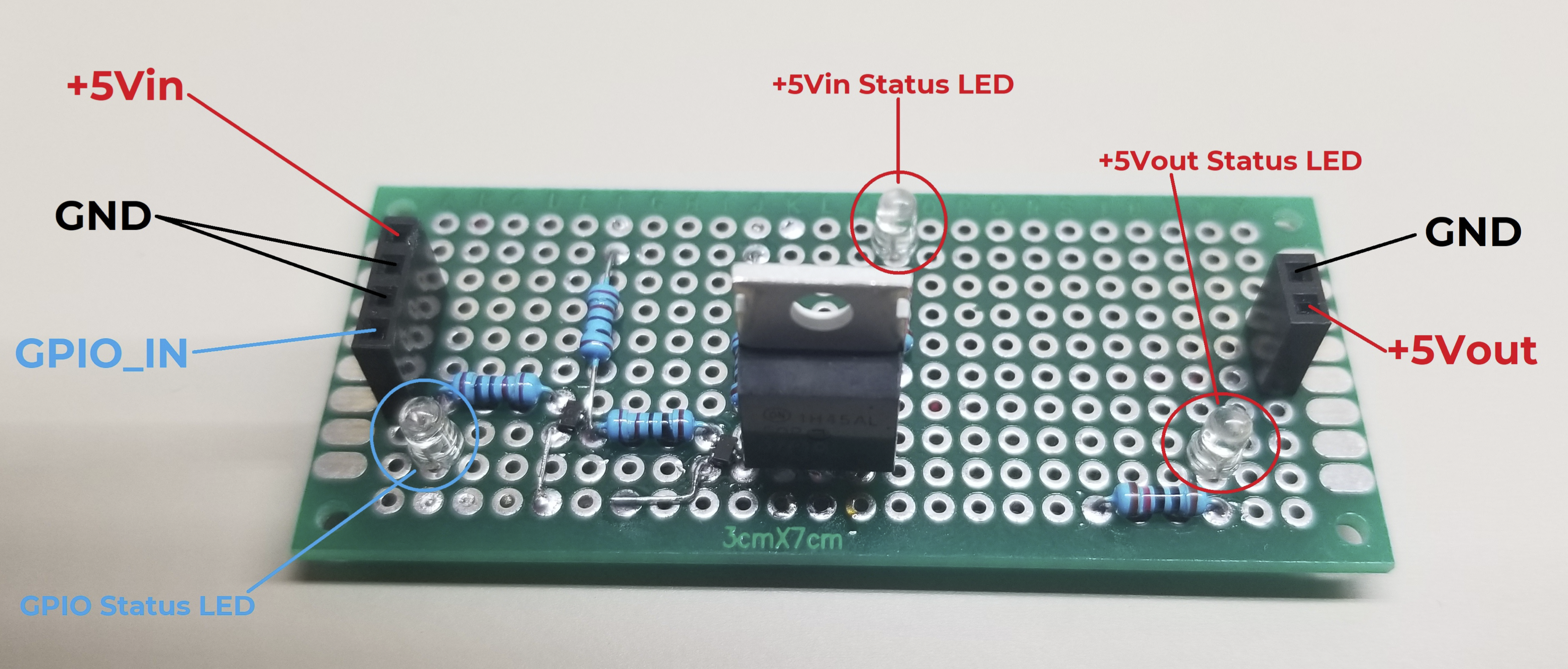}
		
	\caption{Power Control Circuit}
		\label{fig:power_circuit}
\vspace{-5mm}
\end{figure}

The Power Auditor bypasses power to the connected IoT device from an AC adapter. As shown in Figure \ref{fig:power_circuit}, we built a FET-based switch to turn on the connected IoT device via GPIO pin from the Power Auditor device. The current draw of this module is only 20mA when the power output is being provided, while a Power Auditor supplies enough power to most IoT devices. Therefore, IoT devices do not need any modifications in our system.

\begin{figure}[t]
	\centering
	  \subfloat[INA219 Sensor]{%
		\includegraphics[width=0.11\textwidth]{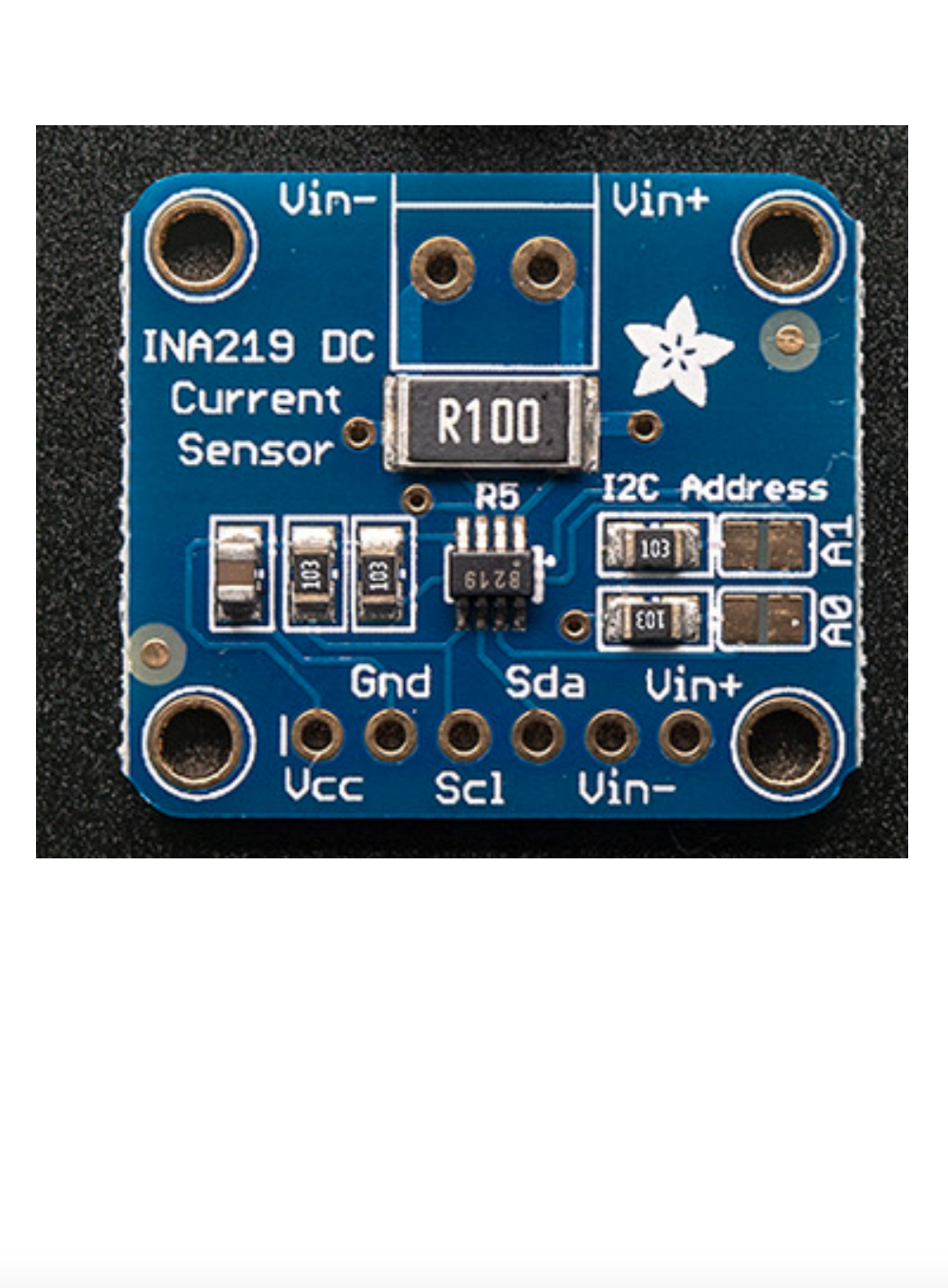}
		\label{fig:sensor}}
	 \subfloat[Logical Circuit of Power Measurement]{%
		\includegraphics[width=0.38\textwidth]{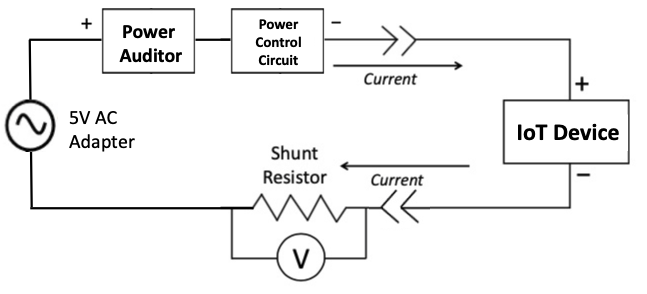}
		\label{fig:current}}
\vspace{-3mm}
	\caption{Power Measurement Design}
	\label{fig:sensingmodule}
\vspace{-3mm}
\end{figure}

We also designed a current circuit for power measurement. To get the current and voltage reading on the connected IoT device, we utilized the current sensor INA219 \cite{ina219}, as displayed in Figure \ref{fig:sensor}. This sensor includes a shunt resistor and provides ADC conversion to the Power Auditor. Figure \ref{fig:current} illustrates how the current sensor provides power consumption data of the IoT device. In this circuit, Power Auditor measures the voltage drop around the shunt resistor at a high frequency. Based on this data, we calculate the current values going through the entire circuit. By doing so, we can measure the power consumption of the IoT device since the voltage input is also fixed. According to the specification of the sensor used, the maximum error rate is approximately 0.5\%, which is negligible.

 After measuring power data, the Power Data Processing module pushes each power reading instance into a local queue. Then, the Networking module fetches the queued data periodically and assembles the collected data to a TCP packet for data transmission to the online classifier. Detailed networking protocol and interface format are illustrated in the next subsection.

\subsection{Sliding Window Protocol} \label{section:4.2}

We introduce the interface design between the Power Auditor and the Data Inferencer for online inference. First, we determined a window size of 1.5 seconds for botnet intrusion input. Jung et al. \cite{Jung-01} revealed that Mirai and its variants have similar time distributions for the initial intrusion during the propagation period, which is less than 1.5 seconds. It should be noted that this invasion time may vary depending on systems or botnets. However, as long as it is noticeable to label, a window size would not be a critical issue.

\begin{figure}[t]
	\centering
	  \subfloat[Client-side Implementation]{%
		\includegraphics[width=0.24\textwidth]{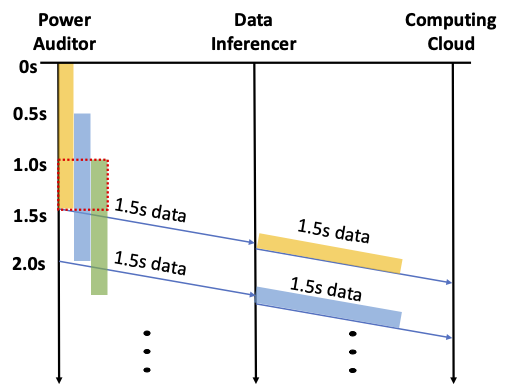}
		\label{fig:net1}}
	 \subfloat[Server-side Implementation]{%
		\includegraphics[width=0.24\textwidth]{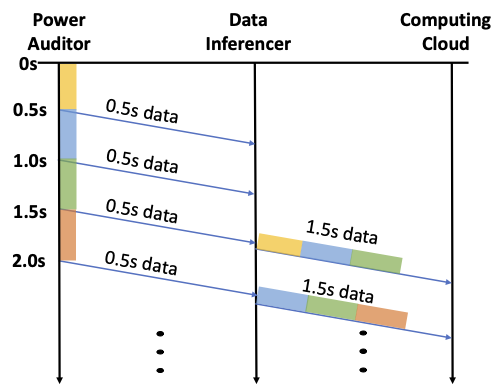}
		\label{fig:net2}}
	\caption{Sliding Window Protocol Options\textemdash\enspace We adopted the Server-side scheme to reduce networking redundancy}
	\label{fig:net}
\end{figure}

\begin{table}[b]
\vspace{-3mm}
	\caption{TCP Packet Interface Format}
	\resizebox{0.48\textwidth}{!}{%
\begin{tabular}{l|l|l|l}
\hline
\textbf{Header} & \textbf{Type} & \textbf{Description} & \textbf{Example} \\ \hline\hline
\textbf{Hostname} & String & Hostname of Power Auditor & smpg1 \\ \hline
\textbf{Message ID} & String & Unique ID for each message & 4da77a50-aeaf-11 \\ \hline
\textbf{Sampling Rate} & Integer & Sampling Rate of Current Sensor (Hz) & 1700 (Hz) \\ \hline
\textbf{Window Size} & Integer & Length of a single window (ms) & 1500 (ms) \\ \hline
\textbf{\begin{tabular}[c]{@{}l@{}}Sliding Window\\ Ratio\end{tabular}} & Integer & Ratio of Overlapping Data & \begin{tabular}[c]{@{}l@{}}1: No Overlapping\\ 2 : 1/2 Overlapping\\ 3 : 1/3 Overlapping\end{tabular} \\ \hline
\textbf{\begin{tabular}[c]{@{}l@{}}Number of\\ Data Points\end{tabular}} & Integer & \begin{tabular}[c]{@{}l@{}}The number of data points \\ in a single TCP packet\end{tabular} & \begin{tabular}[c]{@{}l@{}}850 (Sampling Rate *\\ Window Size / 1000 /\\ Sliding Window Ratio)\end{tabular} \\ \hline
\textbf{Data Points} & Float & List of Power consumption data (mW) & 1854.878, ... (mW) \\ \hline
\end{tabular}}
\label{tab:format}
\end{table}

Next, we applied a sliding window with one-third overlap for better classification accuracy. Our reasoning is that malicious instances can be truncated within a single window. 
Different overlapping ratios are also possible but one needs to consider the trade-off between classification accuracy and network bandwidth. We were able to achieve real-time prediction with the one-third overlapping ratio. Figure \ref{fig:net} illustrates the overlapping sliding window scheme in our DeepAuditor system. As shown in Figure \ref{fig:net1}, if a Power Auditor transmits a data packet of 1.5 seconds every 0.5 seconds, this will create redundant packets. Instead, a Power Auditor reads 0.5 seconds of data and sends it out once collected, as illustrated in Figure \ref{fig:net2}. The Data Inferencer then receives the packet every 0.5 seconds. After receiving three consecutive packets, the Data Inferencer assembles the last three packets and feeds them into the pre-trained CNN classifier for online inference. 
By doing so, we avoided unnecessary network redundancy in a distributed setting.

\begin{figure}[t]
    \centering
      \includegraphics[width=0.48\textwidth]{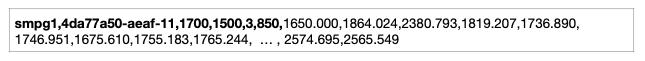}
      \caption{TCP Packet Example in Sliding Window Scheme}
    \label{fig:example}
    \vspace{-5mm}
\end{figure}

We then implemented the interface format accordingly. Table \ref{tab:format} illustrates the packet header format. The number of data points is determined based on the following values. For example, our sampling rate is 1700, and the window size is 1.5 seconds. Consequently, the number of data points in a single instance for classification is $1700 \times 1.5 = 2550$. Since we adopted the server-side sliding window scheme, the Power Auditor also set the Sliding Window Ratio header to 3. Finally, the number of data points in a single TCP packet is $2550 \div 3 = 850$ in our system. The message body contains a list of the power-sensing data. Figure \ref{fig:example} describes an example of the raw TCP packet data.

\section{Distributed CNN Classifier Design} \label{section:5}

In Section \ref{section:5.1}, we present a one-dimensional CNN architecture. Section \ref{section:ppip} then describes a distributed inference protocol between the DeepAuditor components that offloads CNN computations and protects data leakage.

\begin{figure}[b]
	\centering
	\includegraphics[width=0.5\textwidth]{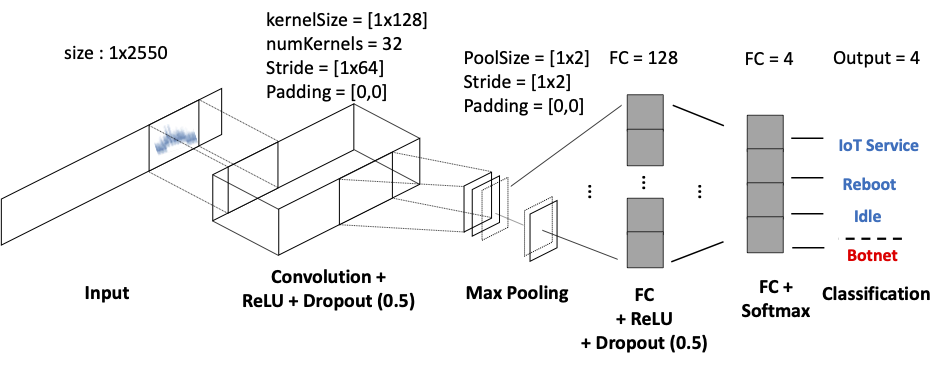}
\vspace{-5mm}
	\caption{1-D CNN Model for the power-trace classification}
	\label{fig:cnn}
\end{figure}

\begin{figure*}[t]
	\centering
	  \subfloat[Dropout Layer]{%
		\includegraphics[width=0.24\textwidth]{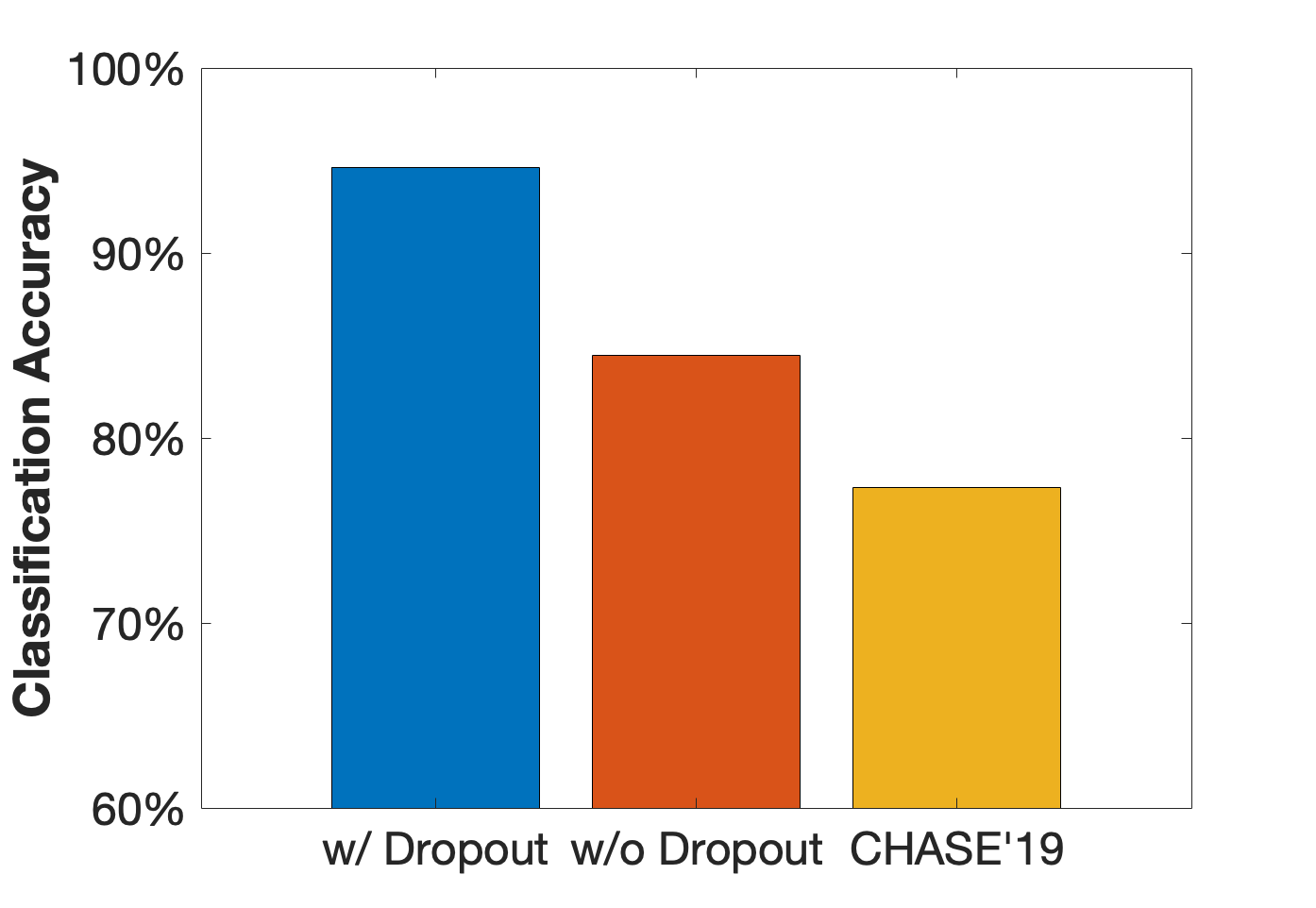}
		\label{fig:impact_dropout}}
	 \subfloat[Pooling Layer]{%
		\includegraphics[width=0.24\textwidth]{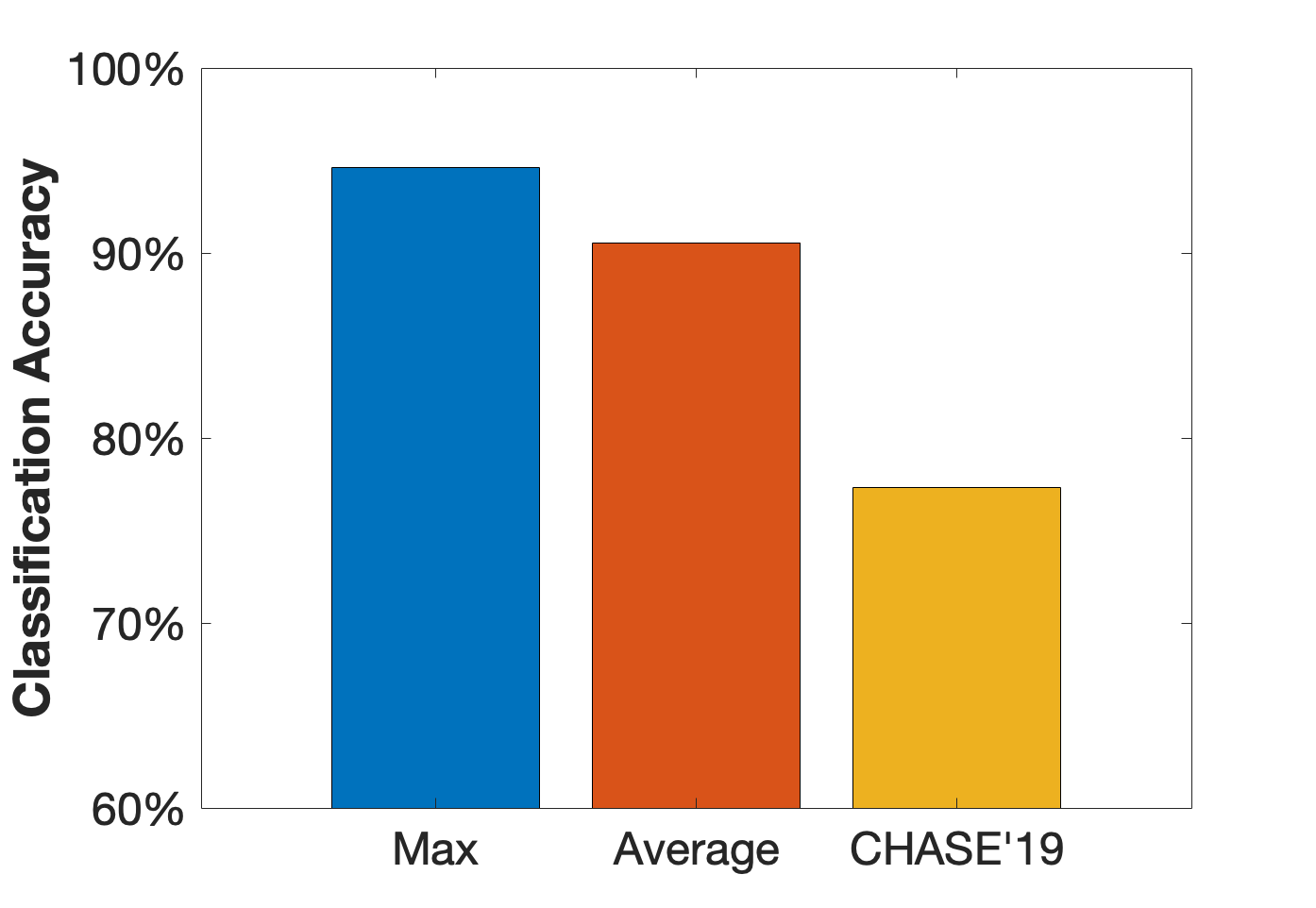}
		\label{fig:impact_pooling}}
	 \subfloat[Convolutional Layer]{%
		\includegraphics[width=0.24\textwidth]{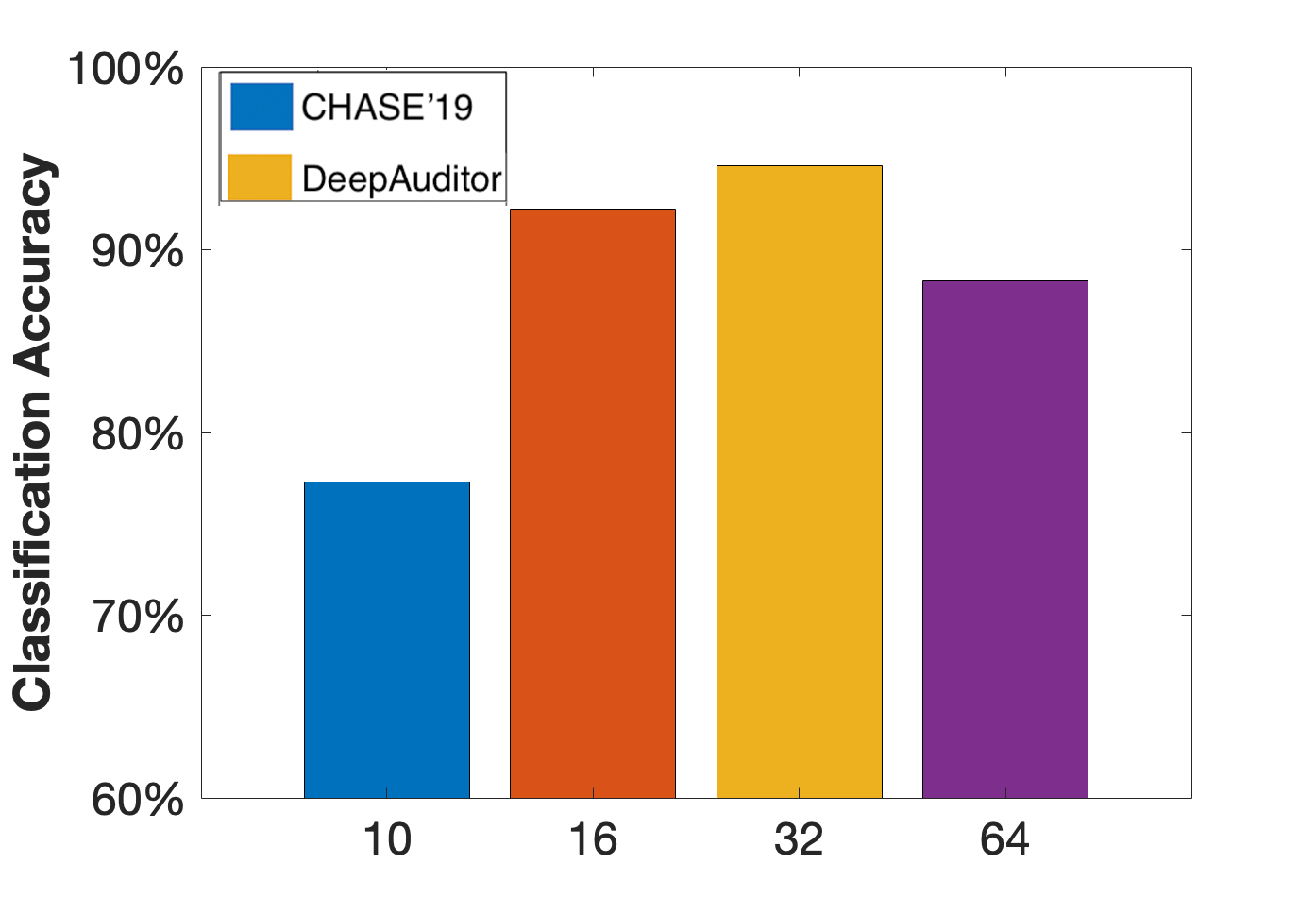}
		\label{fig:impact_kernels}}
	 \subfloat[Fully Connected Layer]{%
		\includegraphics[width=0.24\textwidth]{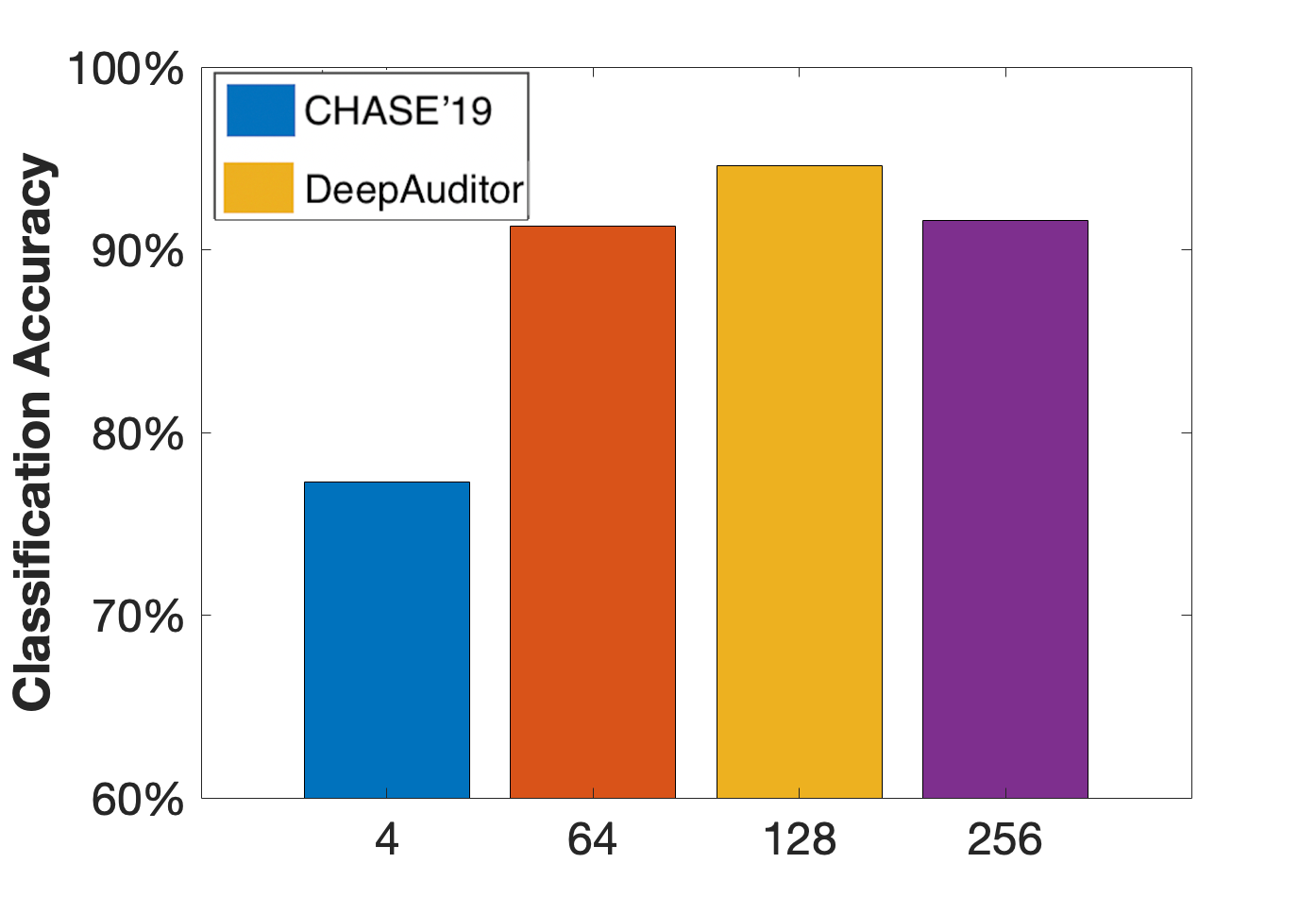}
		\label{fig:impact_fc}}
		
	\caption{Impact of CNN Design Factors}
	\label{fig:impacts}
	\vspace{-3mm}
\end{figure*}

\subsection{1-D CNN Classifier Design} \label{section:5.1}
We designed a 1-D Convolutional Neural Network (CNN) architecture for botnet intrusion detection that outperforms the state-of-the-art model (CHASE'19) \cite{Jung-01}. Jung et al. introduced a CNN model and conducted an offline evaluation. Although this CNN classifier showed promise in identifying subtle differences in power traces, this model performed poorly in some experiments, especially against unseen patterns. For example, the leave-one-out tests introduced a prediction accuracy of below 80\%, which suggests a possible overfitting problem. Indeed, achieving high accuracy in the leave-one-out tests is very important in our system. In a real-world deployment, it is not possible to train all the cases or botnets beforehand. Therefore, the poor leave-one-out results motivated us to propose an enhanced CNN model to avoid overfitting and increase classification accuracy for IoT botnet intrusion.

\subsubsection{CNN Design Philosophy}
Figure \ref{fig:cnn} provides an overview of the proposed 1-D CNN architecture. This CNN classifies time-series power-trace input as belonging to one of four behavior classes in IoT devices: {IoT Service, Idle, Reboot, Botnet Intrusion}. Thus, we seek to distinguish malicious intrusions from other common behavior.

Because the DeepAuditor processes power consumption data, the input layer prepares one-dimensional input to feed into the convolutional layer. The power sensing model has a sampling rate of 1.7kHz and a window length of 1.5 seconds (See Section \ref{section:4.2}). Thus, the input instance size is (1 x 2550). We then use 32 one-dimensional (1 x 128) kernels at a stride size of 64. 
Consequently, the convolutional layer computes a dot product between the power traces and 32 kernels.

Moving forward, we carefully chose the design factors of the 1-D CNN to solve the complex problem of botnet intrusion detection while avoiding overfitting. To demonstrate the effectiveness of each design factor, we used a public dataset, which was collected in 2019, for botnet intrusion detection \cite{misc:iot_dataset} containing power trace data from four different behavior classes (IoT Service, Idle, Reboot, Botnet Intrusion) from three types of IoT devices (Security Camera, Router, Voice Assistant) and three IoT botnets (Mirai, Sora, Masuta). In the dataset, each class has 2,000 instances of 1.5 seconds of power-consumption data. The CHASE'19 CNN classifier performs well in self-evaluation tests where the training and test data are from the same type of device. However, this CNN architecture performs poorly in leave-one-out tests where the test dataset is from a different type of device than the training dataset. We can infer from this that the CNN model is either not powerful enough or overfitted to the specific datasets. 

\begin{table}[b]
\vspace{-3mm}
	\caption{Comparison of Leave-one-out tests}
	\resizebox{0.46\textwidth}{!}{%
\begin{tabular}{c|c|c|c|c}
\hline
\textbf{System} & \textbf{Metric}  & \textbf{Leave-router-out} & \textbf{Leave-voice-assistant-out} & \textbf{Leave-Masuta-out} \\ \hline\hline
\textbf{CHASE'19 \cite{Jung-01}} & Accuracy & 77.3\% & 80.7\% & 79.6\% \\ \hline
\multirow{4}{*}{\textbf{DeepAuditor}} & Accuracy & 94.6\% & 89.1\% & 95.5\% \\ \cline{2-5} 
 & Recall & 98.2\% & 94.7\% & 98.4\% \\ \cline{2-5} 
 & Precision & 94.7\% & 87.4\% & 95.4\% \\ \cline{2-5} 
 & F1-Measure & 96.4\% & 90.9\% & 96.9\% \\ \hline
\end{tabular}
}
\label{tab:loso}
\vspace{-5mm}
\end{table}

\subsubsection{Impact of Design Factors}
In this vein, we compared our design factors with the previous model to show that our model outperforms it. In particular, we used the leave-router-out test to illustrate the performance difference. In this test, the training dataset includes security camera and voice assistant data, while the test dataset has the unseen router power-trace data. Figure \ref{fig:impacts} demonstrates the design factors of our 1-D CNN. First, Figure \ref{fig:impact_dropout} shows that the dropout layer enhanced the validation accuracy from 84\% to 94\%. The dropout layer is helpful because it reduces the number of trainable features to prevent overfitting during training. The CHASE’19 model did not use a dropout layer, so the test and validation accuracy had a gap of 10\% and the validation accuracy was below 80\%. In Figure \ref{fig:impact_pooling}, we observe that the max-pooling method enables to achieve a better classification accuracy than the average-pooling method. This result suggests that the max-pooling method can capture dramatic power-trace spikes well, while the average pooling method smoothes out those power spikes that may include important information. In Figure \ref{fig:impact_kernels}, we achieved a better accuracy by increasing the number of kernels from 10 to 32. Note that using 64 kernels decreased the prediction performance because the unnecessary learning ability (more features) overfitted to the dataset. Lastly, Figure \ref{fig:impact_fc} illustrates that we determined the best number of neurons in the fully connected layer is 128. As illustrated, increasing the number of neurons to 256 hurt the validation accuracy, which hints at a possible overfitting problem.

Overall, we designed a powerful CNN classifier for botnet intrusion detection by optimizing design choices including hyper-parameters. As a result, our CNN classifier outperforms the baseline model, which is too simple and overfitted to the dataset in leave-one-out tests.


\subsubsection{Offline Validation of the Proposed Model}
For practical deployments, neural network models need to be accurate against unseen patterns. Therefore, we further conducted offline experiments to determine the robustness of our model. Table \ref{tab:loso} compares the DeepAuditor and CHASE'19 models.

Leave-one-device-out tests are meaningful because they show the robustness of the classifier in the practical deployment environment. For example, it is not possible to train power-trace data from all IoT devices because there are numerous IoT devices in the market. Thus, our classifier needs to be robust against unseen data from new and future IoT devices. As shown in the previous subsection, in the leave-router-out tests, we trained power data from a security camera and a voice assistant. Then, the router dataset was used to test the classification accuracy. Our DeepAuditor CNN classifier achieved an accuracy of 94.6\%, while the CHASE’19 model predicted power traces with an accuracy of 77.3\%. In the leave-voice-assistant-out test, we also achieved a nearly 10\% improvement in accuracy.

Leave-one-botnet-out tests are also crucial since they demonstrate the CNN's ability to detect new botnet attacks. As discussed in Section \ref{section:2}, Mirai and its variants utilized brute-force attacks, which are still adversaries' preferred options for intrusion into IoT devices. Although these attacks use slightly different procedures, they all follow the model of brute-force attacks followed by post-processing jobs. Thus, it is feasible to identify malicious intrusions of different botnets through power traces. We achieved over 95\% prediction accuracy, which outperforms the baseline classifier's accuracy of below 80\%. The above results clearly show that our model is generalized to different types of devices or botnets.


\subsection{Privacy-preserved Inference Protocol}\label{section:ppip}


Based on our CNN classifier design, we propose a privacy-preserved inference protocol. This protocol was designed between the Data Inferencer and the Computing Cloud to offload computations without data leakage. We will theoretically validate the data protection of the protocol in Section \ref{section:6.3}.

\subsubsection{Motivation and Design Principle}

To offload computation resources, we separate multiplication and summation operations of the convolutional and fully connected layers into the two distributed components. In particular, Data Inferencer only computes cheap plaintext summations, while Computing Cloud carries out the multiplication operations in the ciphertext. 
Simultaneously, to make it more efficient, we encode our input data in both Data Inferencer and Computing Cloud following the convolution kernel’s order rather than keeping the data as original format. By doing this, we also avoid requiring time-consuming permutation operations \cite{GAZELLE} and offline secret sharing strategies in both of the convolutional and fully connected layers. This is an improvement on other works \cite{MiniONN, qiao, delphi}. After that, we feed the output directly into the next layer.

To make it possible, we utilize Packed Homomorphic Encryption (PHE) to allow Data Inferencer to encrypt the power trace data before uploading it to Computing Cloud. Then, the Computing Cloud runs the CNN computations on the encoded ciphertext. Thus, the Data Inferencer encodes multiple plaintext data elements into one ciphertext and high-efficiently carries out element-wise homomorphic computations in a Single Instruction Multiple Data (SIMD) manner \cite{pack}. This tool is particularly useful for our system as our input instance includes thousands of sampling data due to the high sampling rate. Overall, our design uses the CKKS-based PHE that works on element-wise float point data addition and multiplication in ciphertext \cite{ckks, seal}.

\begin{protocol}{Privacy-preserving CNN Inference}
\textit{Inputs:} Received power trace $\mathbf{X}$ 
\label{design protocol}
\\
\textit{Outputs:} Prediction Status $l$ on the given trace $\mathbf{X}$
\sbline
\textit{The Protocol:}
\begin{enumerate} [label=\arabic*.]
\item 
Data Inferencer receives raw data {X} from Power Auditor, encrypts it as $[{X}']_{C}$, and sends $[{X}']_{C}$ to Computing Cloud.
\item
Computing Cloud computes $[U]_{C} = K_{1}W_{1}'[X']_{C} + K_1 B_{1}' + N_{1}$ where $K_{1}$ is a non-zero positive random vector, $W_{1}'$ is the encoded weight, $B_{1}'$ is the encoded bias, $N_{1}$ is a pseudo-random zero-sum vector. Finally, Computing Cloud sends $[U]_{C}$ to Data Inferencer.
\item
Data Inferencer decrypts $[U]_{C}$ and computes $Z_j=\sum_{i=0}^3 U_{i+j}$, where $j$ is the convolution block index. Then, it feeds the result to the ReLU activation function and max pooling layer. Finally, it gets the result ${Y}$.
\item
Data Inferencer encodes and encrypts ${Y}$ as $[{Y'}]_{C}$, and then sends it to Computing Cloud.
\item
Computing Cloud removes the random number $K_1$ and computes the $[V']_{C} = K_{2}W_{2}'[Y']_{C} + K_2 B_{2}' + N_{2}$, where the $K_2$ is a non-zero positive random number, $N_2$ is a pseudo-random zero-sum vector, and the subscript 2 of $W$ and $B$ indicates the corresponding variables at the fully connected layer in our CNN model. Then, Computing Cloud sends $[V']_{C}$ to Data Inferencer.
\item
Data Inferencer decrypts $[V']_{C}$ and performs the summation similarly to Step 3 on each hidden unit block. Finally, it feeds the fully connected layer output into the softmax layer to get the final prediction status $l$ for a given power trace. Depending upon the policy, Data Inferencer takes further actions if $l$ represents a malicious intrusion.
\end{enumerate}
\end{protocol}

\subsubsection{Detailed Procedures}
\begin{figure}[t!]
\centering
\resizebox{0.49\textwidth}{!}{%
\centerline{\includegraphics[width=0.5\textwidth]{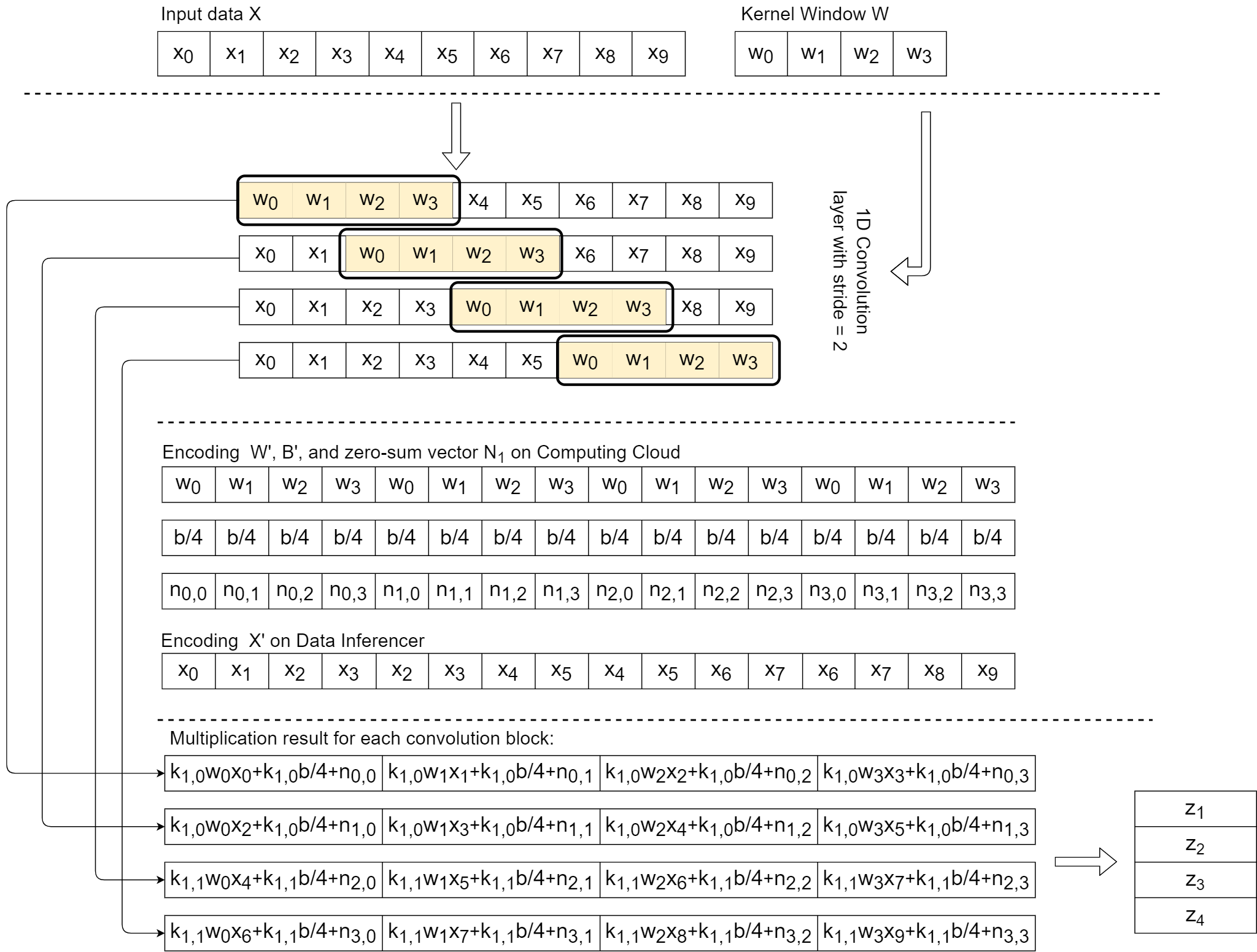}}
}
\caption{Data operations on Data Inferencer and Computing Cloud in Steps 2 and 3} 
\label{fig:ckks}
\vspace{-5mm}
\end{figure}

The detailed privacy-preserved inference protocol is described in Protocol 1 and Figure \ref{fig:ckks}. In order to better explain the key idea, we take an identical CNN model with a smaller size input instead of the original input size. Recall that the input size of our CNN model is (1 x 2550), and the kernel size is (1 x 128) with the stride size 64 (See Section \ref{section:5.1}). Instead, we use a CNN model as an example whose kernel size is (1 x 4) with the stride size 2. 

Let $X$ denote the received raw data by Data Inferencer from Power Auditor. The Data Inferencer first utilizes a PHE package that uses one packed vector to store multiple encrypted plaintext data. In Step 1, to implement the convolution function over the ciphertext, Data Inferencer encodes the data $X$ to $X'$, as illustrated in Figure \ref{fig:ckks}. Correspondingly, Computing Cloud encodes the weight $W_{1}$ and bias $B_{1}$ into packed vectors $W'$ and $B'$ in Step 2. With such encoding, the convolution between $X$ and $W$ can be implemented as the element-wise multiplication between $X'$ and $W'$, plus $B'$.

Steps 2 and 3 show how we securely implement the convolutional layer among ciphertext. After Computing Cloud receives the ciphertext $[X']_{C}$ from Data Inferencer, Computing Cloud uses Eq. (\ref{eq:compute_z}) to compute the homomorphic multiplication result.

\begin{equation}
\label{eq:compute_z}
    [U]_C = K_1 \times W'_1 \times [X']_{C} + K_1 \times B'_1 + N_1,
\end{equation}

The purpose of using random numbers $N_1$ and $K_1$ in Eq. (\ref{eq:compute_z}) is to prevent Data Inferencer from inferring the model parameter $W'_1$ from its received message $[U]_C$. Computing Cloud first generates a zero-sum vector $N_{1} \in \mathbb Z$, which is a vector of pseudo-random numbers such that $N_{1} = \sum_{j=0}^3 n_{i, j}=0$ $(0 \leq j \leq 3)$, to mask each multiplication result, as illustrated in Fig. \ref{fig:ckks}. Then, Computing Cloud multiplies a non-zero positive random number vector $K_{1}=[k_{1,0}, k_{1,1}]$ to mask all multiplication results. With both masks $N_{1}$ and $K_{1}$, Data Inferencer is unable to learn the parameter $W_{1}'$ and $B_{1}'$ based on $[U]_C$ and $X'$. Note that $N_{1}$ is different across different kernels in convolution. Simultaneously, recall that the pool size of max pooling size is 2, $K_{1}$ is different for every two kernels to make sure the correctness for the output of max pooling layer. 

Steps 5 and 6 are similar to Steps 2 and 3 but implement the fully connected layer. However, Computing Cloud only requires to choose a non-zero positive random number $K_{2}$ in Step 5 to mask the ciphertext multiplication result. At the end of Step 6, Data Inferencer directly feeds the weighted sum result of the fully connected layer into the softmax layer to infer the Power Auditor status $l$. If the $l$ value represents a malicious intrusion on the IoT Device, Data Inferencer can take further actions, such as sending a notification to the administrator or shutting down the IoT device. 

\section{Online System Implementation}\label{section:implementation}
For performance evaluation, we have implemented a prototype system in a distributed setting. Section \ref{section:7.1} describes the prototype implementation and Section \ref{section:datacollection} shows the new dataset we collected.

\subsection{DeepAuditor Implementation} \label{section:7.1}
We have built a prototype DeepAuditor for proof of concept. Table \ref{tab:testbed} describes our testbed environment. First, Raspberry Pi 3 devices serve as both Power Auditors and IoT devices for prototyping purposes. We also used a desktop in the same local network for Data Inferencer. Computing Cloud was deployed in a department cloud server outside the local network. 

In Power Auditors, we implemented the proposed modules in Python for real-time data collection. For prototyping IoT devices, we installed an open-source camera software called MotionEye \cite{etc:motion} on the connected IoT devices. This software includes a motion detection feature as well as a video streaming feature. Thus, we consider the Raspberry Pi 3 device running MotionEye as an IoT device. Likewise, we also installed the Google AIY project \cite{etc:aiy} on another Raspberry Pi 3 device and conducted voice commands on it. The Power Auditor is connected to an AC adapter. The IoT device is then supplied with power through the Power Auditor, as shown in Figure \ref{fig:auditor}. 

Next, we implemented the sliding window scheme between the Power Auditor and the Data Inferencer. Since the Power Auditor sends segmented power traces of 1.5-second data every 0.5 seconds, the Data Inferencer is required to predict the power trace data every 0.5 seconds.

\begin{table}[t]
\caption{Online System Testbed}
\resizebox{0.49\textwidth}{!}{%
\begin{tabular}{c|c|c|c|c|c}
\hline
 & \textbf{Power Auditor} & \textbf{IoT Device} & \textbf{Mirai Bot} & \textbf{\begin{tabular}[c]{@{}c@{}}Data\\Inferencer\end{tabular}} & \textbf{\begin{tabular}[c]{@{}c@{}}Computing\\Cloud\end{tabular}} \\ \hline\hline
\textbf{\begin{tabular}[c]{@{}c@{}}Network\\ Deployment\end{tabular}} &  \multicolumn{4}{c|}{\begin{tabular}[c]{@{}c@{}}Local Network\end{tabular}} &  Cloud\\ \hline
\textbf{\begin{tabular}[c]{@{}c@{}}Hardware\\ Platform\end{tabular}} & Raspberry Pi 3 & Raspberry Pi 3 & Raspberry Pi 3 & \begin{tabular}[c]{@{}c@{}}Apple Mac\\ Mini Desktop\end{tabular} &  Linux Server\\ \hline
\textbf{\begin{tabular}[c]{@{}c@{}}CPU\\ \end{tabular}} & \begin{tabular}[c]{@{}c@{}}1.4GHz,\\ Quad-core\end{tabular}& \begin{tabular}[c]{@{}c@{}}1.4GHz,\\ Quad-core\end{tabular}& \begin{tabular}[c]{@{}c@{}}1.4GHz,\\ Quad-core\end{tabular}& \begin{tabular}[c]{@{}c@{}}2.5GHz,\\ Dual-core\end{tabular}& \begin{tabular}[c]{@{}c@{}}2.6GHz,\\ 2*16 cores\end{tabular} \\ \hline
\textbf{\begin{tabular}[c]{@{}c@{}}Memory\\ \end{tabular}} & 1GB&1GB&1GB& 8GB& 250GB \\ \hline
\textbf{\begin{tabular}[c]{@{}c@{}}Sensor\\ used\end{tabular}} & \begin{tabular}[c]{@{}c@{}}Current Sensor\\ (INA219A)\end{tabular} & \begin{tabular}[c]{@{}c@{}}- Camera Sensor\\ - External Microphone\end{tabular} & N/A & N/A & N/A \\ \hline
\textbf{\begin{tabular}[c]{@{}c@{}}Operating\\ System\end{tabular}} & Raspbian 9 & Raspbian 10 & Raspbian 9 & 
\begin{tabular}[c]{@{}c@{}}macOS 10.15\end{tabular}
 & 
\begin{tabular}[c]{@{}c@{}}Ubuntu 18.04\end{tabular}
 \\ \hline
\textbf{\begin{tabular}[c]{@{}c@{}}Software\\ Module\end{tabular}} & \multicolumn{1}{l|}{\begin{tabular}[c]{@{}l@{}}- Power Monitoring\\ - Networking\\ - Power Controlling\end{tabular}} & \begin{tabular}[c]{@{}c@{}}- Video Streaming\\ - Motion Detection\\- Voice Assistant\end{tabular} & \begin{tabular}[c]{@{}c@{}}Mirai \\ (Scanner/Loader)\end{tabular} & \begin{tabular}[c]{@{}c@{}} Privacy-preserved \\ Online Classifier \end{tabular} & \begin{tabular}[c]{@{}c@{}} Privacy-preserved \\ Online Classifier \end{tabular} \\ \hline
\end{tabular}}
\label{tab:testbed}
\vspace{-3mm}
\end{table}

As mentioned in Section \ref{section:ppip}, our privacy-preserved protocol utilized CKKS-based PHE \cite{ckks}, which involves three parameters \footnotemark: 1) polynomial modulus degree $N$, 2) ciphertext scale $s$, and 3) coefficient modulus. In our protocol, we set up the CKKS-related encryption parameters as follows: 1) Parameters for PHE scheme are selected for a 128-bit security level, 2) the selection of polynomial modulus degree $N$ is a smaller degree that allows to encode $N/2$ elements into one ciphertext. In our case, we selected $N = 32768$ for the convolution layer and $N = 4096$ for the fully connected layer, 3) a ciphertext scale $s=2^{40}$ is enough to store all the intermediate results in the convolutional layer, and a ciphertext scale $s=2^{20}$ is enough to store all the intermediate results in the fully connected layer, and 4) In seal \cite{seal}, the modulus switching chain is set up as coefficient modulus for ciphertext against exponential noise growth during ciphertext operations. In our case, we selected the coefficient modulus $(60, 40, 40, 60)$ for the convolutional layer and $(30, 20, 20, 30)$ for the fully connected layer.

\footnotetext{The reader is referred to \cite{ckks, seal} for more details.}

With the above parameters, we fulfilled the CNN classifier in the Data Inferencer and the Computing Cloud. According to Protocol 1, the classification procedures comprise six steps; steps 1, 3, 4, and 6 are implemented in the Data Inferencer, whereas steps 2 and 5 are implemented in the Computing cloud. Step 6 makes a final decision for power-trace prediction.

To satisfy the real-time prediction requirement, we adopted pipeline processing \cite{pipeline} in the six inference steps; each step is being executed in parallel in the Data Inferencer and the Computing Cloud. In other words, the Data Inferencer and the Computing Cloud do not have to wait until all the steps are completed. This pipelining increases the throughput of the instructions, which eventually leads to our real-time inference for multiple IoT devices simultaneously. We will discuss the performance in Section \ref{section:7.5}.

\begin{table}[t]
\centering
	\caption{The Collected Dataset of Power Traces }
	\resizebox{0.38\textwidth}{!}{%
\begin{tabular}{cc|c|c|c}
\hline
\multicolumn{2}{c|}{\textbf{Class}} & \textbf{Description} & \textbf{\begin{tabular}[c]{@{}c@{}}Number of\\ Instances\end{tabular}} & \textbf{\begin{tabular}[c]{@{}c@{}}Total number\\ of Instances\end{tabular}} \\ \hline\hline
\multicolumn{2}{c|}{Idle} & \begin{tabular}[c]{@{}c@{}}When IoT Service\\ is not running\end{tabular} & 4693 & 4693 \\ \hline
 \multicolumn{2}{c|}{  \begin{tabular}[c]{@{}c@{}}IoT Device\\ Service\end{tabular} } & Security Camera \cite{etc:motion} & 5976 & \multirow{2}{*}{8176} \\ \cline{3-4}
 &  & Voice Assistant \cite{etc:aiy} & 2200 &  \\ \hline
\multicolumn{2}{c|}{Reboot} & \begin{tabular}[c]{@{}c@{}}When system\\ is rebooting\end{tabular} & 2200 & 2200 \\ \hline
 \multicolumn{2}{c|}{  \begin{tabular}[c]{@{}c@{}}Botnet\\ (Mirai) \cite{etc:mirai}\end{tabular} }& \begin{tabular}[c]{@{}c@{}}Malicious behavior\\ while system is Idle\end{tabular} & 2000 & \multirow{2}{*}{3000} \\ \cline{3-4}
\multicolumn{2}{c|}{} & \begin{tabular}[c]{@{}c@{}}Malicious behavior\\ while system is running\\ IoT service\end{tabular} & 1000 &  \\ \hline
\end{tabular}
}
\label{tab:dataset}
\vspace{-3mm}
\end{table}

\subsection{Dataset Collection for Online Test} \label{section:datacollection}
As discussed earlier, our classifier predicts which of the four classes a power instance belongs to. Even though the previous section already demonstrated the robustness of our classifier design on the public dataset, we newly collected power traces from two different types of IoT devices to demonstrate the real-time performance. Thus, we created a new dataset in our testbed environment, as shown in Table \ref{tab:dataset}. 

Table \ref{tab:dataset} summarizes the collected dataset. In our environment, we generated a specific scenario and collected over 2,000 power instances for each class. For example, the data we collected for the Idle class comprises 4,693 instances of 1.5-second power traces when the IoT service was not running. We also collected power traces when the IoT service was running or the IoT device was rebooting. For example, we used the open-source MotionEye \cite{etc:motion} for security cameras and the Google AIY project \cite{etc:aiy} for voice assistant prototypes. For the Botnet class, we downloaded an open-source code of Mirai from Github \cite{etc:mirai} and built it on an IoT bot testbed. To generate Mirai instances in our local network, we modified the source code to attack only our IoT devices. Then, we collected 3,000 instances of malicious attacks when the IoT service is running or the system is idle.

\section{Online System Evaluation} \label{section:7}
In Section \ref{section:7.2}, we validate the performance of the Power Auditor device. Section \ref{section:7.3} then demonstrates the system-level online classification performance. In Section \ref{section:6.3}, we theoretically analyze the data protection of the privacy-preserved inference protocol. Finally, Section \ref{section:7.5} illustrates the scalability evaluation of the system.

\subsection{Power Auditor Performance} \label{section:7.2}

In Table \ref{tab:device}, we compare our power auditing device with the off-the-shelf device Monsoon Power Monitor \cite{msoon} that was used in the offline study \cite{Jung-01}. Our Power Auditor supports a sampling rate of up to 1.7kHz and a voltage of up to 5.5V. These ranges are less than the Monsoon device and thus may be limited in measuring the power of large appliances with built-in computation units, e.g., smart fridges or smart microwaves. However, it is still enough to audit most IoT devices' power consumption. On the contrary, the Power Auditor is much smaller and lighter than the Monsoon device, which makes our device convenient for ubiquitous power measurement. More importantly, the proposed device is capable of measuring power-trace data in real-time for online inference. The Power Auditor also supplies power to connected devices while measuring power consumption. 

\begin{table}[t]
\caption{Power Auditing Devices Comparison}
\centering
\resizebox{0.31\textwidth}{!}{%
\begin{tabular}{c|c|c}
\hline
 &
  \textbf{\begin{tabular}[c]{@{}c@{}}Our\\ Power Auditor\end{tabular}} &
  \textbf{\begin{tabular}[c]{@{}c@{}}Monsoon\\ Power Monitor \cite{msoon}\end{tabular}} \\ \hline\hline
\textbf{\begin{tabular}[c]{@{}c@{}}Sampling\\ Rate\end{tabular}} &
  Up to 1.7kHz &
  Up to 5kHz \\ \hline
\textbf{\begin{tabular}[c]{@{}c@{}}Measurement\\ Range\end{tabular}} &
  \begin{tabular}[c]{@{}c@{}}Up to 5.5V\\ Up to 2.3A\end{tabular} &
  \begin{tabular}[c]{@{}c@{}}Up to 13.5V\\ Up to 6A\end{tabular} \\ \hline
\textbf{Dimension} &
  3" x 2" x 1" &
  8" x 6" x 3" \\ \hline
\textbf{Weight} &
  0.3lb &
  4lb \\ \hline
\textbf{Software} &
  \begin{tabular}[c]{@{}c@{}}Text-based\\ Python Software\end{tabular} &
  \begin{tabular}[c]{@{}c@{}}Window GUI\\ Software\end{tabular} \\ \hline
\textbf{\begin{tabular}[c]{@{}c@{}}Online\\ Measurement\end{tabular}} &
  Available &
  Not Applicable \\ \hline
\textbf{Price} &
  \$25 &
  \$929 \\ \hline
\end{tabular}
}
\vspace{-3mm}
\label{tab:device}
\end{table}

\begin{table}[b]
\centering
\caption{Online System Performance per IoT Device}
\resizebox{0.48\textwidth}{!}{%
\begin{threeparttable}

\begin{tabular}{c|c|c|c|c|c}
\hline
&\textbf{\begin{tabular}[c]{@{}c@{}}CPU Load\\ (Max\tnote{1}\enspace)\end{tabular}} & \textbf{\begin{tabular}[c]{@{}c@{}}Memory\\ Usage\end{tabular}} & \textbf{\begin{tabular}[c]{@{}c@{}}Network\\ Bandwidth\end{tabular}} & \textbf{\begin{tabular}[c]{@{}c@{}}Processing\\ Delay\end{tabular}} &
\textbf{\begin{tabular}[c]{@{}c@{}}Power\\Consumption\end{tabular}} \\ \hline\hline
\textbf{Power Auditor} & 76\% (400\%) & 10MB & 120Kbps & 25ms & 2\si{\watt} (400\si{\mA}) \\ \hline
\textbf{Data Inferencer} & 40\% (200\%) & 50MB & 11.375Kbps & 160ms & \textemdash \\ \hline
\textbf{Computing Cloud} & 35\% (3200\%) & 30MB & \textemdash & 360ms &\textemdash \\ \hline
\end{tabular}
  \begin{tablenotes}
    \item[1] Maximum CPU Load depends on the number of CPU cores, e.g., Dual-core has a maximum 200\% CPU load.
  \end{tablenotes}
\end{threeparttable}
}
\label{tab:metrics}
\vspace{-5mm}
\end{table}

Table \ref{tab:metrics} illustrates the performance metrics for a single Power Auditor working with our servers. In short, the resources of Raspberry Pi 3 are more than capable of supporting the Power Auditor device. For example, the maximum CPU load in the Power Auditor is up to 76\%, which used roughly 20\% of the Raspberry Pi's quad-core computing power. Memory (RAM) usage during online auditing is only 10MBytes out of Raspberry Pi's 1GB (1\%). Based on the proposed sliding window design, the required network bandwidth between the Power Auditor and the Data Inferencer is only 120Kbps. This bandwidth is extremely small and therefore can be covered by Bluetooth or even the Zigbee protocol. Moreover, the power consumption of the Power Auditor is approximately 2\si{\watt} during the real-time inference. That is about an extra \$2 in cost per year for single-device monitoring, which is similar to the power consumption of existing smart plug devices \cite{smartplug_power}. Note that these results were the same regardless of IoT device type. Due to space constraints, we did not include those results. Overall, our results confirm that the Power Auditor is lightweight enough for an online auditing device, and we plan to build a prototype upon real-world demonstration. 

\subsection{Online CNN Classifier Performance} \label{section:7.3}

We measured online classification results in a laboratory setting. 
We provide the classification accuracy and other metrics associated with this test to validate the classifier performance. We generated power instances of each class in real-time and conducted online inference to obtain metrics. For example, the IoT device was in idle status for the Idle class, while the IoT device streamed video or conducted voice commands for the IoT Service class. For the Mirai intrusion class, a bot device sneaked into the IoT device, and we measured the inference results on the intrusion events.

\begin{figure}[t]
	\centering
    \includegraphics[width=0.38\textwidth]{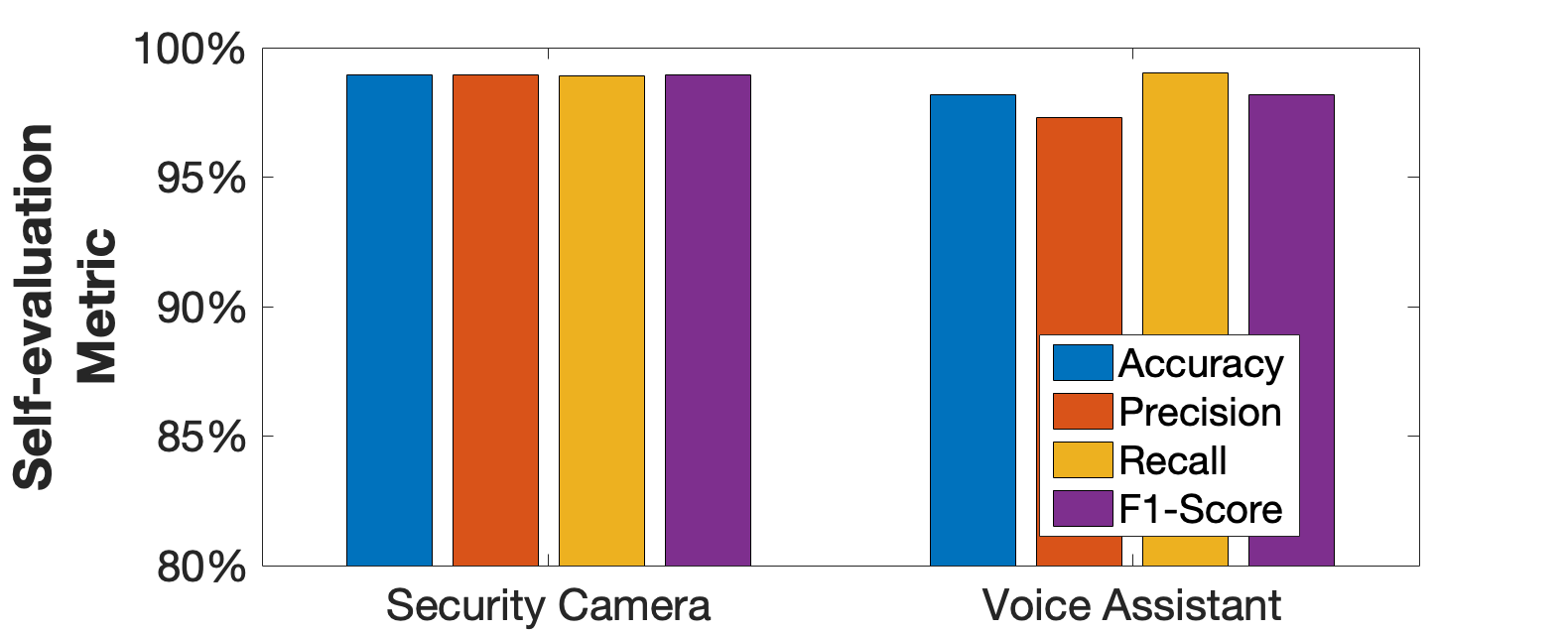}
	\caption{Online Classification Results}
	\label{fig:online_class}
\vspace{-5mm}
\end{figure}

\begin{figure}[b]
	\vspace{-5mm}
	\centering
	  \subfloat[Processing Time per Protocol Step ]{%
		\includegraphics[width=0.25\textwidth]{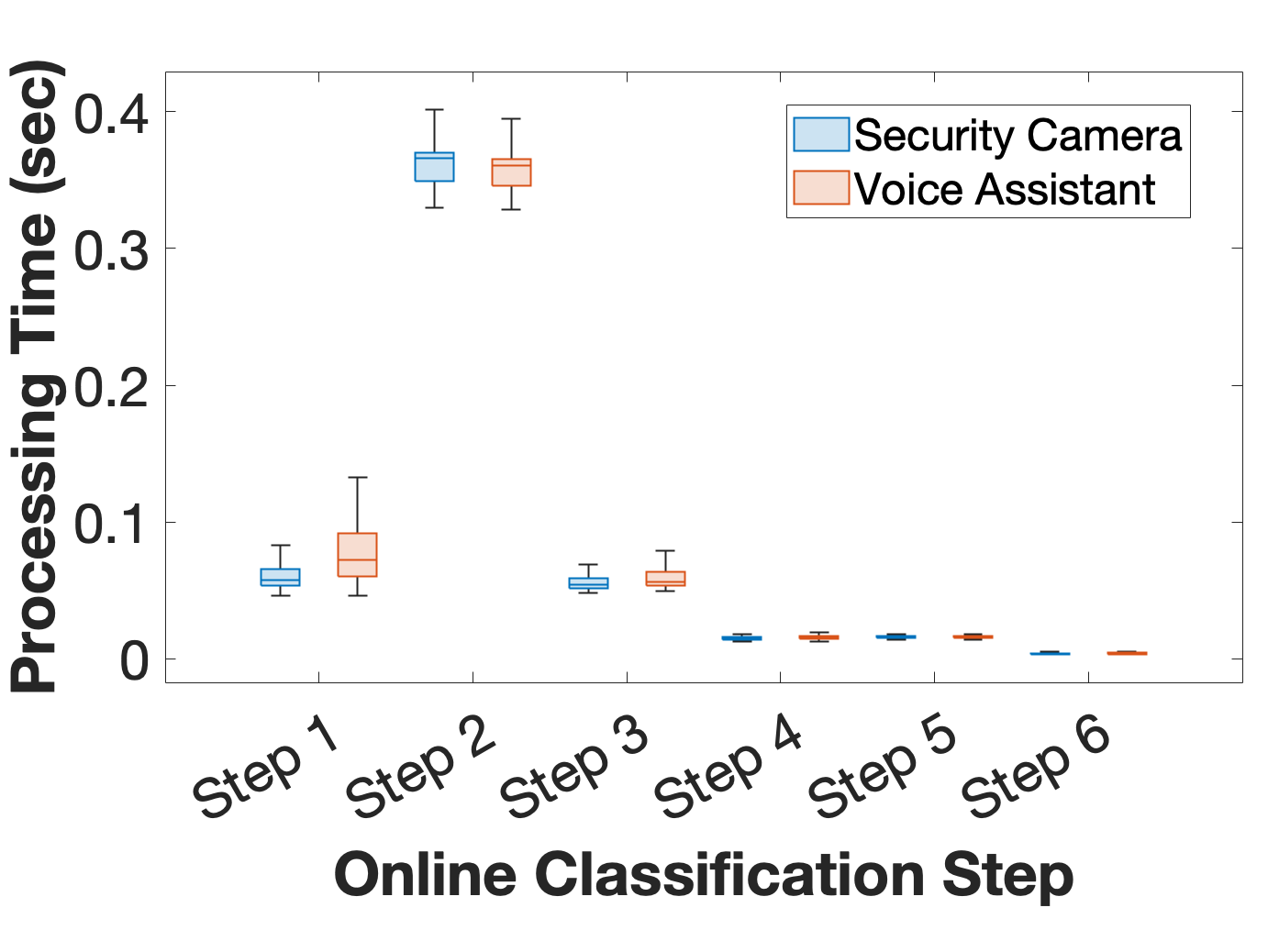}
		\label{fig:dist}}
	 \subfloat[Empirical CDF of Response Time]{%
		\includegraphics[width=0.25\textwidth]{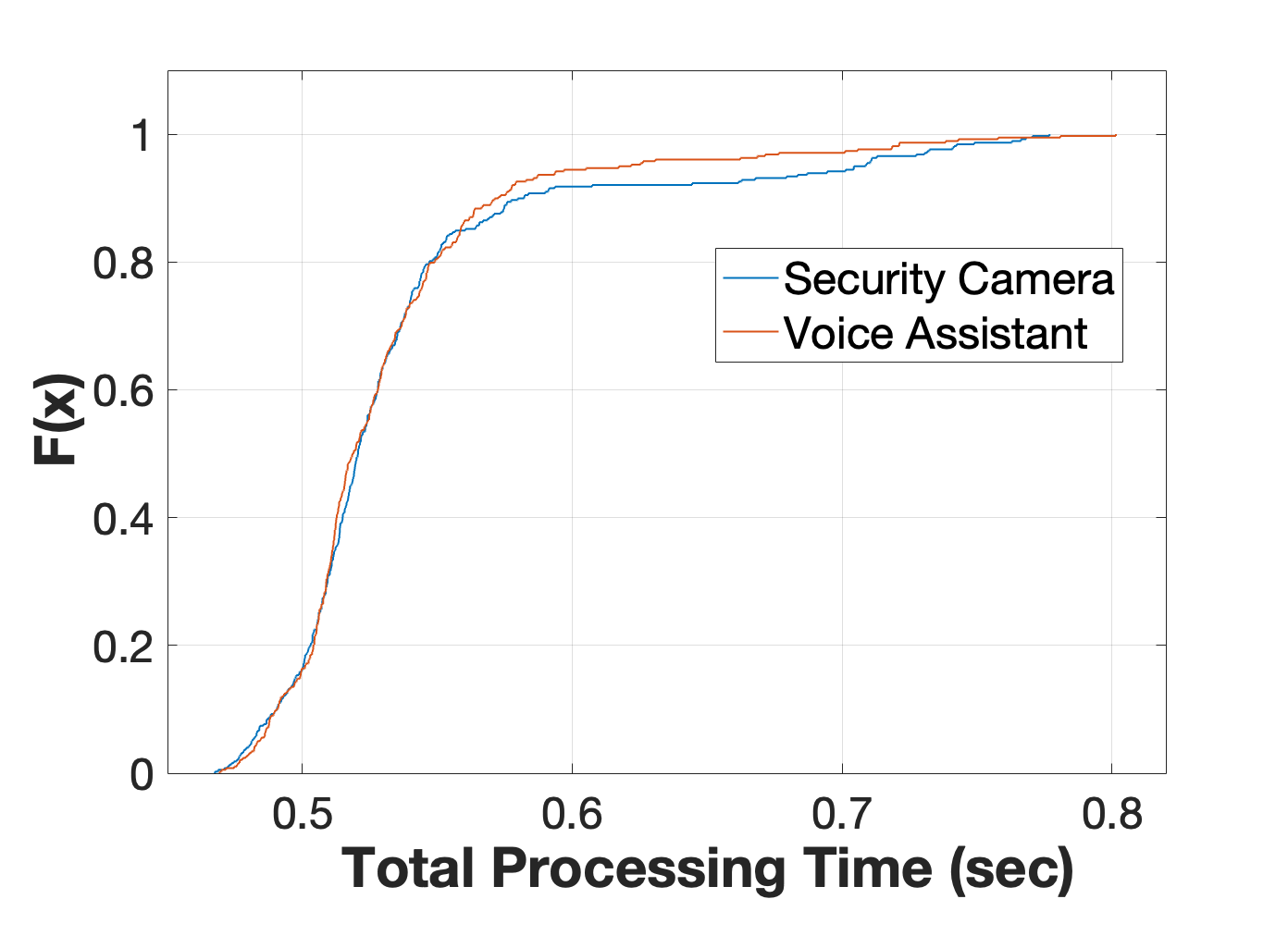}
		\label{fig:cdf}}
	\caption{Online Classifier Processing Time}
	\label{fig:oc}
	\vspace{-3mm}
\end{figure}

Table \ref{tab:metrics} shows the performance of each distributed component as the classifier is deployed in separate servers. These metrics were obtained when the DeepAuditor processed continuous data from a single Power Auditor monitoring an IoT device. The total delay per power instance is approximately 520 milliseconds: $160ms$ in the Data Inferencer and $360ms$ in the Computing Cloud. In our testbed, the transmission delay between the Data Inferencer and the Computing Cloud is approximately $20ms$, while the local network delay is less than 1ms. The network bandwidth between the Data Inferencer and the Computing Cloud is 11.375Kbps, and the memory usage is 50MBytes in the Data Inferencer and 30MBytes in the Computing Cloud. Overall, these numbers are not overwhelming for online inference since we utilized cloud resources. 

Figure \ref{fig:online_class} also shows the online classification results for each IoT device. The results demonstrate the exceptional classification ability of the CNN classifier. For example, we achieved an overall accuracy of 98.95\%. The Precision and Recall metric values for both tests are also above 98\%. Thus, the distributed classifier is able to distinguish different patterns of device behavior in real-time, as trained.


Furthermore, Figure \ref{fig:dist} illustrates the processing delay of each step in the privacy-preserved inference protocol. For instance, it takes about $350ms$ to complete the Convolution procedure (Step 2) on Computing Cloud regardless of device type, which is the majority of the entire online classification. Other steps consume relatively fewer computing resources. In addition, Figure \ref{fig:cdf} demonstrates an empirical CDF function of the online classification response time. For both devices, we observe that over 80\% of the inferences were done in $550ms$ or less. Note that our system is required to classify real-time instances per Power Auditor every $500ms$. To maximize the process throughput, we applied pipeline processing to the inference protocol. Thus, the entire performance mainly rests on step 2, which is the most time-consuming job in our classifier. Even though most instances are completed about $550ms$  after the Data Inferencer receives the power instance, the Computing Cloud is able to classify two instances per second securely because step 2 takes at most $400ms$, as presented in Figure \ref{fig:dist}. Overall, the DeepAuditor as an entire system is able to reliably predict input instances every $500ms$ in real-time.

\subsection{Theoretical Analysis of the Privacy-preserved Inference Protocol} \label{section:6.3}
In this subsection, we demonstrate that our distributed classifier design is secure in that 1) Computing Cloud cannot obtain the client's power-trace data, and 2) Data Inferencer cannot obtain the model parameters $W$ and $B$ of the CNN model in Computing Cloud. Hence, there is no information leakage between Computing Cloud and Data Inferencer.

We used a security analysis method called \emph{ideal/real world} paradigm \cite{Goldreich}. Let P1 denote Data Inferencer with input $x$ and P2 Computing Cloud with input $y$. Let $f = (f_1, f_2)$ be a set of functionality and $\pi$ be a protocol which is implemented in our online system for computing $f$. The party $P_i$ wishes to obtain the protocol output $f_i(x, y)$ $(i \in \{1, 2\})$. The view of $P_i$ during an execution of $\pi$ on the set of input $\Bar{x} = \{x, y \}$ is denoted as $\mathit{view}_i^{\pi}(\Bar{x})$, and it equals to $(w, r_i;m_i^1, \dots, m_i^t)$ where $w \in {x, y}$ is the input of $P_i$ for $i \in \{1, 2\}$. $r_i$ is equals to the set of random numbers inside $P_i$, and $m_i^j$ represents the $j$-the message received by $P_i$.

The adversary can compromise any one party in our model, but not the majority and all of them are non-colluded. There exists a probabilistic polynomial-time simulator that has same functionality as $\pi$. The security of $\pi$ is defined as follows:

A Protocol $\pi$ can securely execute $f$ in the presence of
semi-honest adversaries, if for any exists a probabilistic polynomial-time simulator $S_i$, such that for the corrupted parities $P_i$, it has: 
\begin{gather}
\{(S_1(1^n, x, f_1(\Bar{x})), f(\Bar{x}))\} \equiv \mathit{view}_i^{\pi}(\Bar{x})\} \nonumber\\
\{(S_2(1^n, y, f_2(\Bar{x})), f(\Bar{x}))\} \equiv \mathit{view}_i^{\pi}(\Bar{x})\} 
\end{gather}

where $\equiv$ represents computationally indistinguishable, and $i \in \{1, 2\}$.

The protocol $\pi$ is secure against semi-honest adversaries if the views of the real-world execution are computationally indistinguishable from the view of the simulator in ideal world. In the following section, we use the method above to prove our online system is secure against semi-honest adversaries.


\subsubsection{\textit{Data Security against Compromised Data Inferencer}}
We assume that Data Inferencer is compromised by Adversary A. We use the simulator \textsf{sim} to behave as Adversary A which interacts with \textsf{f}. The \textsf{sim}, \textsf{f}, and Data Inferencer conduct the following steps:

1) \textsf{sim} first gets $[X']_{C}$ from Data Inferencer. Then, it sends $[\Bar{X}']_{C}$ to \textsf{f}. \textsf{f} returns $[U]_C$ to \textsf{sim}. 

2) Starting Data Inferencer, \textsf{sim} generates random number $\Bar{W}$, $\Bar{B}$, $\Bar{K}$ and $\Bar{N}$, and computes $[U']_C=\Bar{K}\Bar{W}[X']_{C} +\Bar{K}\Bar{B}+\Bar{N}$ by Eq. (\ref{eq:compute_z}). Then, \textsf{sim} sends $[U']_C$ to Data Inferencer.

3) \textsf{sim} decrypts and outputs $U$ and $U'$

The above protocol is secure against adversary Data Inferencer based on the randomness of $K'$ and $N'$ which makes the view of $U'$ that is generated by \textsf{sim} computationally indistinguishable from the view of $U$ that the real output from \textsf{f}. This conclusion can also be applied to the view of the Fc layer in our system.

\subsubsection{\textit{Data Security against Compromised Computing Cloud}} 
Similarly, we assume that Computing Cloud is compromised by Adversary A which interacts with \textsf{sim}. The \textsf{sim}, \textsf{f}, and Computing Cloud conduct the following steps:

1) \textsf{sim} first gets $W_1$, $B_1$, $K_1$, $N_1$, $W_2$, $B_2$, $K_2$, and $N_2$ from Computing Cloud. Then, it sends $W_1$, $B_1$, $K_1$, $N_1$, $W_2$, $B_2$, $K_2$, and $N_2$ to \textsf{f}. \textsf{f} returns None to \textit{sim}. 

2) Starting Computing Cloud, \textsf{sim} generates and encrypts a group of random numbers $\Bar{X}$ as $[\Bar{X}]_C$. Then, \textsf{sim} sends it to Computing Cloud. 

3) \textsf{sim} receives $[U']_C$ from Computing Cloud, \textsf{sim} generates and encrypts a group of random numbers $\Bar{Y}$ as $[\Bar{Y}]_C$.  Then, \textsf{sim} sends $[\Bar{Y}]_C$ to Computing Cloud.

4) \textsf{sim} outputs $([\Bar{X}]_C, [\Bar{Y}]_C)$.

Our system is secure against Computing Cloud because the view of the Data Inferencer's input data $[X]_C$ and the intermediate output $[Y']_C$ is computationally indistinguishable from the $[\Bar{X}]_{C}$ and $[\Bar{Y}]_C$ that is generated by \textsf{sim} based on the fact that the PHE algorithm is semantically secure \cite{pack}.

\begin{table}[b]
\centering
\vspace{-3mm}
\caption{Comparison of Computation Complexity} 
\label{complexity:computiaon}
\resizebox{0.48\textwidth}{!}{%
\begin{tabular}{c|c|c|c}
\hline
\textbf{Methodology} & \textbf{Permutation} & \textbf{Multiplication} & \textbf{Addition} \\ \hline\hline
\textbf{Gazelle-Convolution} & $\mathcal{O}(r)$ & $\mathcal{O}(rC)$ & $\mathcal{O}(rC)$ \\ \hline
\textbf{DeepAuditor-Convolution} & $0$ & $\mathcal{O}(C)$ & $\mathcal{O}(C)$ \\ \hline
\textbf{Gazelle-FC} & $\mathcal{O}(k)$ & $\mathcal{O}(k\frac{Cn_in_o}{n})$ & $\mathcal{O}(k\frac{Cn_in_o}{n})$ \\ \hline
\textbf{DeepAuditor-FC} & $0$ & $\mathcal{O}(\frac{Cn_in_o}{n})$ & $\mathcal{O}(\frac{Cn_in_o}{n})$ \\ \hline
\end{tabular}}
\label{tab:result}
\vspace{-3mm}
\end{table}

\subsubsection{Computation Complexity Analysis} \label{section:7.3.3}

We analyzed the overall computation complexity for the convolutional layer and fully connected (FC) layer of DeepAuditor in Table \ref{complexity:computiaon}. Let denote that $r$ is kernel size for the convolutional layer and $C$ is the number of output channels for the convolutional layer. We compared our work with a naive method of Gazelle \cite{GAZELLE}, which is a more state-of-the-art protocol with speed-up than some classic privacy-preserving inference protocol like \cite{cryptonets}. Based on our benchmark on Protocol 1, Computing Cloud needs $C$ times ciphertext multiplication and addition to compute the intermediate result $[U]_C$. Meanwhile, Data Inferencer only conducts a cheap plaintext summation $[U]_C$ to complete the convolution output $Z_j$. The total computation cost for our protocol design is much less than the Gazelle, which requires $r$ times ciphertext permutation, $rC$ times ciphertext multiplication, and $rC$ times ciphertext addition.

To improve the computation efficiency in the FC layer. We extended our data encode operation presented in Figure \ref{fig:ckks}, which enables all max-pooling outputs to be packed into one ciphertext. Let $n_i$ denote the number of data in each output channel of max-pooling, $n_o$ denote the output data for the FC layer, $n$ denote the number of slots for one ciphertext, and $k$ denote the kernel size for the FC layer. Each ciphertext can hold $\frac{Cn_in_o}{n}$ data. Compared with Gazelle's input packing method \cite{GAZELLE}, the Computing Cloud in our protocol needs $k$ times less than ciphertext multiplication and addition. The FC layer of our protocol also does not require ciphertext permutation, while Gazelle still requires $k$ times ciphertext permutation.

\subsection{Scalability Evaluation} \label{section:7.5}

As shown in Figure \ref{fig:2}, a single Computing Cloud supports multiple Power Auditors. We deployed the online CNN classifier in a distributed environment to offload computation resources and handle multiple IoT devices simultaneously. In our testbed, we evaluated how many IoT devices the cloud servers could support for intrusion detection.

\begin{figure}[t]
	\centering
	\vspace{-3mm}
	  \subfloat[CPU Utilization]{%
		\includegraphics[width=0.25\textwidth]{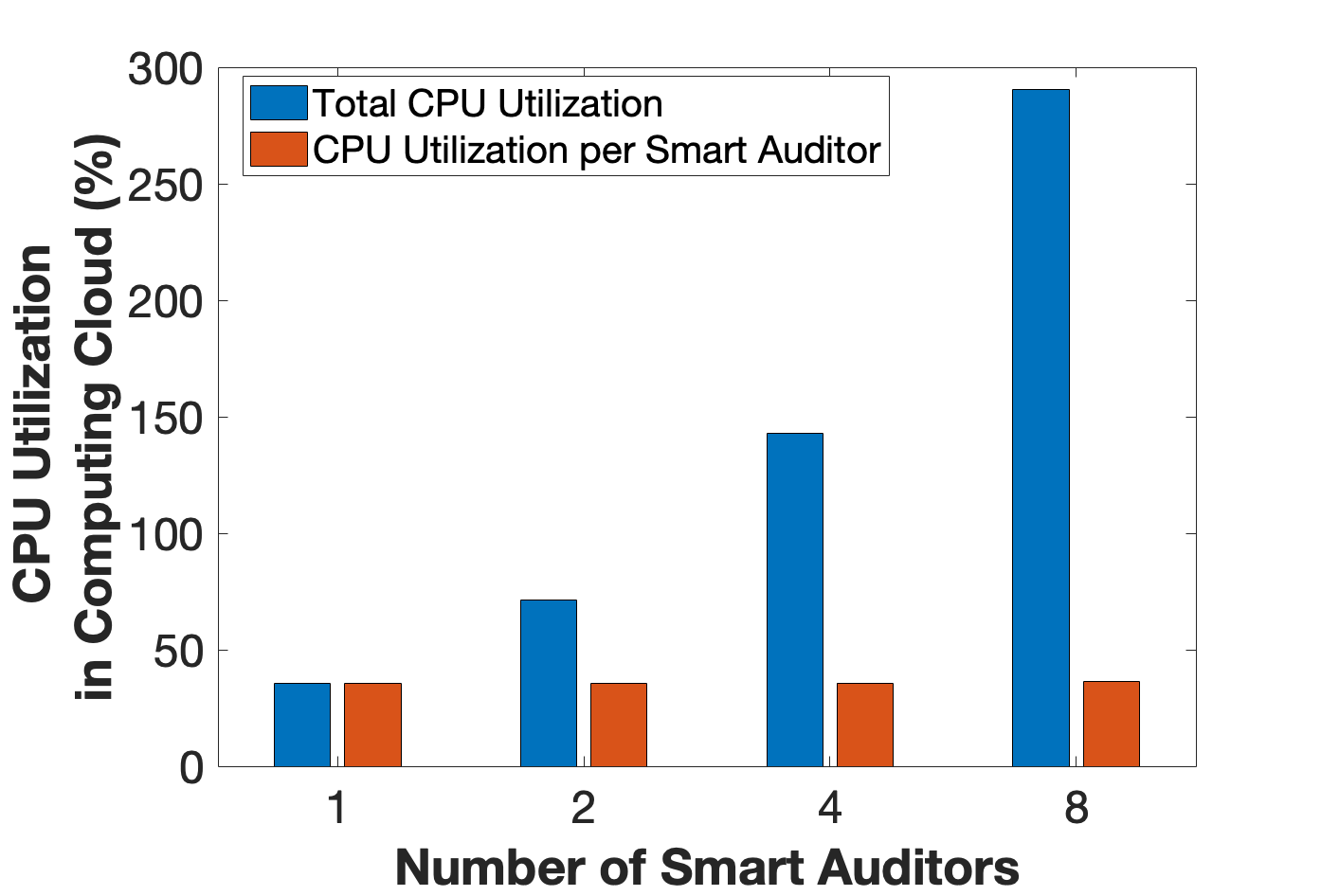}
		\label{fig:cpu}}
	 \subfloat[Processing Time in Step 2 ]{%
		\includegraphics[width=0.25\textwidth]{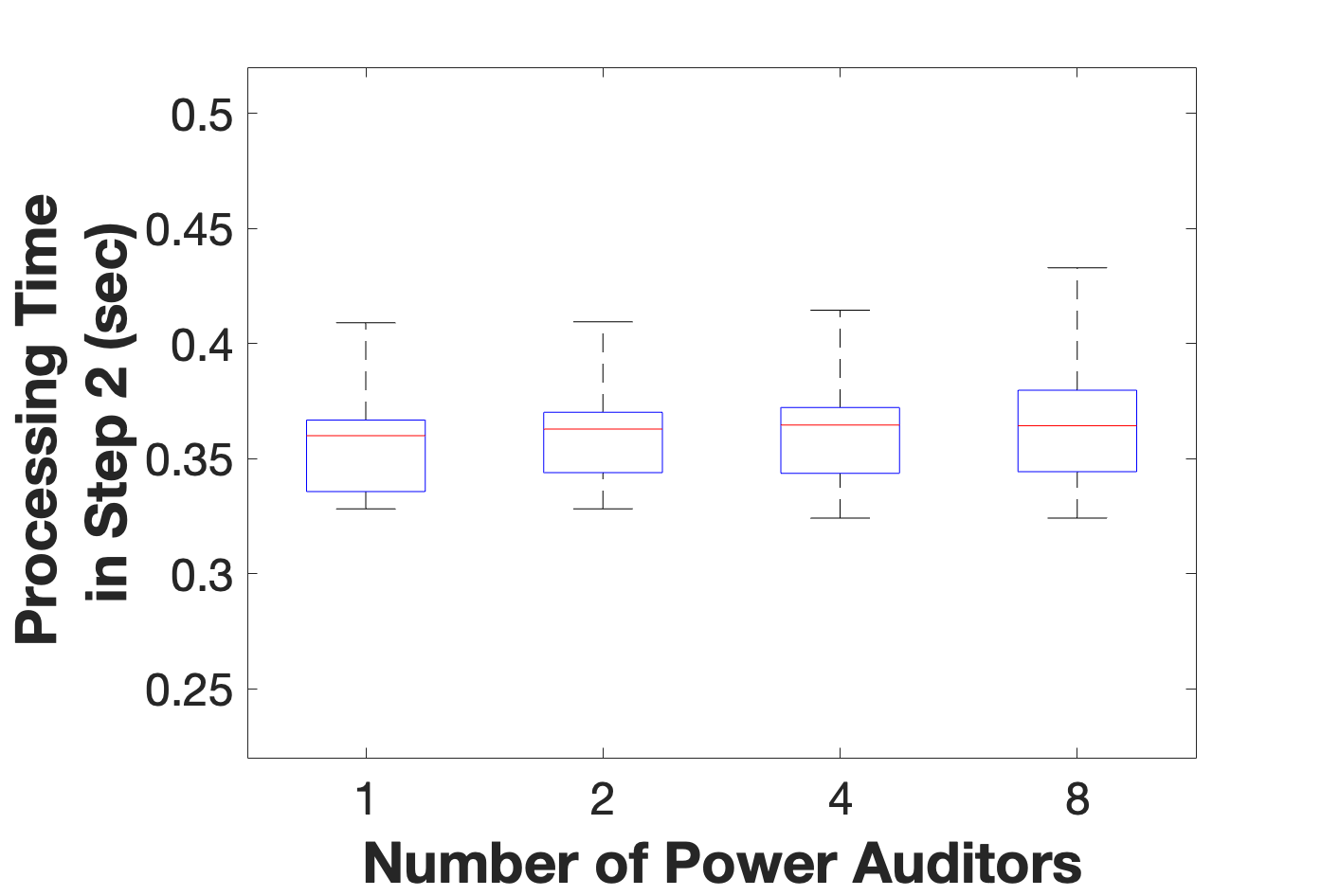}
		\label{fig:scale}}
	\caption{Scalability Performance Evaluation}
	\label{fig:scalability}
	\vspace{-5mm}
\end{figure}

To test scalability, we set up a distributed environment with eight Power Auditors, two Data Inferencers, and one Computing Cloud. Four Power Auditors are connected to each Data Inferencer, both of which are connected to the same Computing Cloud. The eight Power Auditors collect power traces from their respective IoT devices in real-time. Note that system scalability relies on efficient use of available resources, such as throughput and CPU utilization \cite{862209}, regardless of device or botnet type. Figure \ref{fig:scalability} deficits performance results of four tests of 1, 2, 4, and 8 Power Auditors. In Figure \ref{fig:cpu}, as the number of Power Auditors increases, the CPU utilization of the Computing Cloud increases linearly. The Computing Cloud's CPU utilization per Power Auditor is approximately 35\%, which is consistent with the result in Table \ref{tab:metrics}. Moreover, when we tested with eight Power Auditors, the total CPU utilization was less than 300\% out of 3,200\% (32 Cores).

Furthermore, Figure \ref{fig:scale} demonstrates that the DeepAuditor classifies multiple IoT devices' data in real-time. As discussed in Section \ref{section:7.3}, the inference time mostly relies on the processing time in step 2 of our protocol. In Figure \ref{fig:scale}, as we increased the number of Power Auditors, the average processing time does not change substantially. As long as the processing time in step 2 is less than 0.5 seconds, our DeepAuditor system can guarantee real-time inference for online detection. 

Overall, these results demonstrate that the Computing Cloud supports inference computations for multiple IoT devices to the extent that CPU cores are available. For example, step 2 takes approximately $350ms$ on average, and the CPU utilization of Computing Cloud is about 35\% for handling a single Power Auditor. Thus, a single core of Computing Cloud can handle up to three Power Auditors per second. Since the Computing Cloud in our environment has 32 CPU cores, this server can support almost a hundred Power Auditors. We further plan to enhance the convolution processing time in the future because the current performance is bounded by the processing time in step 2.

\section{Discussion} \label{section:8}

In this section, we present our thoughts with regard to limitations and future work.

\subsection{Power Auditor Prototype}
Since we utilized a Raspberry Pi device for prototyping the Power Auditor, our current version is still bulky and costly for ubiquitous power measurement. Several researchers have already developed a small form-factor power meter for communication purposes \cite{jiang2009design} \cite{debruin2015powerblade}. However, because it was not designed for IoT devices, the power ranges are slightly higher than those of many IoT devices, whereas our current Power Auditor is able to monitor low-powered IoT devices. Thus, we plan to make our Power Auditor an AC-plug meter device that can be attached to off-the-shelf IoT devices. This task requires PCB board manufacturing and is shown to be feasible by other works. If so, the Power Auditor will be even smaller and cheaper than the current prototype. 
Moreover, since the smart plug industry has grown dramatically, we believe that in the future, our model can be integrated into generic smart plug devices for online intrusion detection.

\subsection{Real-world Deployment}
DeepAuditor is the first distributed online system to assess the power consumption of IoT botnet intrusion following the emergence of Mirai and similar IoT botnets. Fundamentally, our distributed system components process the power consumption of the connected IoT devices in real-time for intrusion detection, which has never been tackled before. Thus, we focused on demonstrating the validity of our system design since other approaches may not be directly comparable with the DeepAuditor. Nevertheless, our proposed CNN classifier was tested with two different datasets. Table \ref{tab:env} shows the environments of the two datasets, including location, time, and power monitor devices. As demonstrated in the previous sections, our classifier performed well in both environments.

However, our system still needs to be tested in a real-world deployment. Based on the proven concept, we plan to further expand our current system to a wild setting in which commercial IoT devices will be used for validation purposes. We will tackle this task in the future.

\begin{table}[t]
\caption{Our classifier was validated on different datasets}
\resizebox{0.49\textwidth}{!}{%
\begin{tabular}{c|c|c|c|c|c}
\hline
 & \textbf{\begin{tabular}[c]{@{}c@{}}Experimental\\ Environment\end{tabular}} & \textbf{\begin{tabular}[c]{@{}c@{}}Power Monitoring \\ Device\end{tabular}} & \textbf{\begin{tabular}[c]{@{}c@{}}Year \\ Collected\end{tabular}}  & \textbf{\begin{tabular}[c]{@{}c@{}}Number of \\ Device Type\end{tabular}} & \textbf{\begin{tabular}[c]{@{}c@{}}Number of \\ Instances\end{tabular}} \\ \hline\hline
\textbf{CHASE'19 \cite{misc:iot_dataset}} & Lab & \begin{tabular}[c]{@{}c@{}}Monsoon \\ Power Monitor\end{tabular} & 2019 & 3 & 8000 \\ \hline
\textbf{DeepAuditor} & At-home & \begin{tabular}[c]{@{}c@{}}Our Proposed \\ Device\end{tabular} & 2021 & 2 & 15869 \\ \hline
\end{tabular}}
\label{tab:env}
\vspace{-5mm}
\end{table}



 
\section{Related Work}\label{section:9}
Section \ref{section:9.3} presents side-channel studies on IoT security. In Section \ref{section:9.1}, we summarize the related work with regard to IoT security via power auditing. Then, Section \ref{section:9.2} introduces the existing work concerning about the preservation of data privacy.

\subsection{IoT Security via Side-channel Information}\label{section:9.3}

The literature has scrutinized IoT security against botnets for decades \cite{Aversano:9999bh, Anonymous:2021gb}. This area includes network-based solutions as well as power-auditing-based detection methods. While network-based solutions have struggled to address the endpoint security on IoT devices \cite{8629941}, several works have utilized side-channel information, such as electromagnetic (EM) data. Especially for resource-constrained devices, using side-channel information is efficient because it utilizes existing resources instead of requiring many modifications \cite{or2019dynamic}.

Nazari et al. \cite{nazari2017eddie} proposed an EM-based spike detection method to identify injected codes in program execution. The EM spectrum was monitored in order to detect malware, and the results showed promise in using side-channel information.
Sehatbakhsh et al. \cite{sehatbakhsh2018syndrome} also utilized EM-based monitoring to identify anomalous behavior during execution on medical devices.
While EM-based side-channel monitoring also showed potential in program execution, system-level monitoring is preferred in the identification of malicious IoT botnets. This is because IoT botnets often enter and compromise entire target devices \cite{Jung-01}.

\subsection{IoT Security via Power Auditing} \label{section:9.1}

Power side-channel information has also been utilized to infer malicious behavior on end devices. For example, some pioneering works used power side-channel data to detect malign behavior on mobile devices in the early 2010s \cite{Kim:2008cl}\cite{Yang2017OnIB}. Recently, several researchers \cite{Li:ej, Myridakis:2021tg, Jung-01} have worked on IoT devices to characterize malicious behavior as IoT botnets have been popular. Myridakis et al. \cite{Myridakis:2021tg} implemented a power monitoring circuit for botnet prevention for IoT devices. However, this system mainly focused on detecting massive DoS attacks on IoT devices with a spike detection method instead of intrusion detection. Li et al. \cite{Li:ej} addressed energy auditing for physical and cyber attacks. For cyber attacks, it utilized dual-CNNs to infer massive DoS attacks such as network flood using energy meters \cite{Hindle:ug}. Clark et al. \cite{clark2013wattsupdoc} aimed to identify malware behavior on medical devices via power auditing. Similarly, they conducted offline classification experiments on a dataset collected from a single device. Jung et al. \cite{Jung-01} pioneered IoT botnet intrusion detection via power modeling. While their CNN classifier showed promise in detecting subtle differences in power traces, the study was conducted offline with a bulky and expensive power monitor. Thus, it is not practical for ubiquitous botnet detection on IoT devices. In summary, Table \ref{tab:related} summarizes the related works on IoT security via power auditing. Overall, there is still a gap between power auditing techniques and practical IoT intrusion detection. For efficient IoT botnet detection, a scalable real-time solution via power auditing is needed. We are the first to realize a distributed online classification system for botnet intrusion detection on multiple IoT devices via ubiquitous power auditing. 

\begin{table}[t]
	\caption{Comparison of IoT Security via Power Auditing}
	\resizebox{0.49\textwidth}{!}{%
\begin{tabular}{c||c|c|c|c|c|c}
\hline
\textbf{} & 
\textbf{\begin{tabular}[c]{@{}c@{}}Target\\Attacks\end{tabular}} &
\textbf{\begin{tabular}[c]{@{}c@{}}Testbed\\Environment\end{tabular}} &
\textbf{\begin{tabular}[c]{@{}c@{}}Learning\\Method\end{tabular}} &
\textbf{\begin{tabular}[c]{@{}c@{}}Concurrent\\Capacity\end{tabular}} &
\textbf{\begin{tabular}[c]{@{}c@{}}Auditing\\Device\end{tabular}} &
\textbf{\begin{tabular}[c]{@{}c@{}}Classification\\Accuracy\end{tabular}} \\ \hline\hline
\textbf{Jung \cite{Jung-01}} &\begin{tabular}[c]{@{}c@{}}Botnet\\Intrusion\end{tabular} 
&\begin{tabular}[c]{@{}c@{}}Offline\\Modeling\end{tabular} 
&\begin{tabular}[c]{@{}c@{}}1-D\\CNN\end{tabular}
&\begin{tabular}[c]{@{}c@{}}Single\\Device\end{tabular} 
& Monsoon \cite{msoon}
& 96.5\% 
\\ \hline
\textbf{Li \cite{Li:ej}} 
&\begin{tabular}[c]{@{}c@{}}DoS\\Attacks\end{tabular} 
&\begin{tabular}[c]{@{}c@{}}Online\\Classification\end{tabular} 
&\begin{tabular}[c]{@{}c@{}}Dual\\1-D CNNs\end{tabular} 
&\begin{tabular}[c]{@{}c@{}}Single\\Device\end{tabular} 
&\begin{tabular}[c]{@{}c@{}}IoT\\Hardware\end{tabular} \cite{Hindle:ug}
& MSE 0.032 \\ \hline
\textbf{Myridakis \cite{Myridakis:2021tg}} 
&\begin{tabular}[c]{@{}c@{}}DoS\\Attacks\end{tabular} 
&\begin{tabular}[c]{@{}c@{}}Online\\Classification\end{tabular}
&\begin{tabular}[c]{@{}c@{}}Spike\\Detection\end{tabular}
&\begin{tabular}[c]{@{}c@{}}Single\\Device\end{tabular} 
&\begin{tabular}[c]{@{}c@{}}IoT\\Hardware\end{tabular} 
& 100\%
\\ \hline
\textbf{Clark \cite{clark2013wattsupdoc}}
&\begin{tabular}[c]{@{}c@{}}Malware\\Attacks\end{tabular} 
&\begin{tabular}[c]{@{}c@{}}Offline\\Classification\end{tabular}
&\begin{tabular}[c]{@{}c@{}}kNN, RF,\\ Perceptron\end{tabular}
&\begin{tabular}[c]{@{}c@{}}Single\\Device\end{tabular} 
&\begin{tabular}[c]{@{}c@{}}AC\\Outlet\end{tabular} \cite{clark2013security}
& 94\%

\\ \hline
\textbf{DeepAuditor} 
&\begin{tabular}[c]{@{}c@{}}\textbf{Botnet}\\Intrusion\end{tabular} 
&\begin{tabular}[c]{@{}c@{}}\textbf{Online}\\Classification\end{tabular}
&\begin{tabular}[c]{@{}c@{}}\textbf{Distributed}\\1-D CNN\end{tabular}
&\begin{tabular}[c]{@{}c@{}}\textbf{90+}\\Devices\end{tabular} 
&\begin{tabular}[c]{@{}c@{}}\textbf{IoT}\\Hardware\end{tabular} 
&98.9\%\\ \hline
\end{tabular}}
\label{tab:related}
\vspace{-3mm}
\end{table}

\subsection{Preservation of Data Privacy} \label{section:9.2}

Preservation of data privacy has been widely studied in the literature. There are three major approaches. The first approach is differential privacy \cite{dp}, which injects noise into query results, such as perturbating stochastic gradient descent (SGD) \cite{Abadi2016ccs}. However, the additive noise may degrade model accuracy. 
The second approach designs privacy-preserved protocols based on secure multi-party computations. They usually distribute secrets among a group of parties to achieve security computations at the expense of high computational overhead and strong security assumptions \cite{secret1}\cite{secret2}\cite{secret3}. Thus, they are rarely adopted in general scenarios. 

A new solution for privacy preservation was introduced by using the fully homomorphic encryption \cite{gentry2009}. It allows users to encrypt data and offload the computation to a cloud. The cloud computes encrypted data and sends back encrypted results \cite{cryptonets}\cite{MiniONN}\cite{GAZELLE}. However, the nonlinear activation computation cannot be supported by the homomorphic encryption, and the approximation often has to be used. Compared with existing work, our solution is novel in that our proposed scheme capitalizes on the proposed CNN model structure to adopt a smart design to address this problem. 

\section{Conclusion } \label{section:10}
In this paper, we proposed a distributed online intrusion detection system for IoT devices via power auditing. We first developed a portable power-auditing device to measure power side-channel information of IoT devices in real-time. The one-dimensional CNN classifier was then designed and deployed in a distributed setting. The online CNN classifier predicted IoT devices’ behavior with up to 98.9\% accuracy, which outperforms the baseline classifier, especially in leave-one-out tests. In addition to the system components, we also designed distributed protocols to avoid data leakage and reduce networking redundancy. Finally, we evaluated the scalability of the system in a laboratory setting. Altogether, the DeepAuditor system is the first online intrusion detection system that classifies multiple IoT devices’ behavior via power traces. This kind of cloud system that uses power auditing of multiple IoT devices has not been addressed before in the literature, so such system can be used on IoT devices for other purposes.

In the future, we plan to enhance the performance of the inference protocol. Currently, the convolutional layer consumes the majority of the entire processing time. If we reduce that procedure, our system will be more reliable and scalable. In addition to the pre-trained classifier, we further plan to apply unsupervised learning so that users can use their dataset without labeling. This can expedite system deployment in a practical setting.

\section*{Acknowledgements}
This research is partially supported by COVA CCI Cybersecurity Research and Innovation Funding, COVA CCI Cybersecurity Innovation Bridge Fund (Grant \#HC-4Q21-005), COVA CCI Dissertation Fellowship, and NSF grant CNS-2120279.
We would like to thank all the anonymous reviewers for their valuable comments.

\bibliographystyle{abbrv}
\bibliography{acmart, Xin,reference}

\begin{thebibliography}{10}

\bibitem{misc:iot_dataset}
Iot-botnet-detection via power consumption modeling.
\newblock \url{https://woossup.github.io/IoT-Botnet-Detection}.

\bibitem{smartplug_survey}
Smart plug market - growth, trends, covid-19 impact, and forecasts (2021 -
  2026).
\newblock
  \url{https://www.mordorintelligence.com/industry-reports/smart-plug-market}.

\bibitem{smartplug_power}
What is a smart plug and how it eliminates energy waste.
\newblock
  \url{https://www.atlanticenergyco.com/post/smart-plug-energy-savings/}.

\bibitem{Abadi2016ccs}
M.~Abadi, A.~Chu, I.~Goodfellow, H.~B. McMahan, I.~Mironov, K.~Talwar, and
  L.~Zhang.
\newblock Deep learning with differential privacy.
\newblock In {\em Proceedings of the 2016 ACM SIGSAC Conference on Computer and
  Communications Security}, CCS '16, pages 308--318, New York, NY, USA, 2016.
  ACM.

\bibitem{ina219}
Adafruit.
\newblock {\em INA219 Current Sensor}, 2020.
\newblock \url{https://learn.adafruit.com/
  adafruit-ina219-current-sensor-breakout}.

\bibitem{Anonymous:dcm0bUHc}
M.~Antonakakis.
\newblock {Understanding the Mirai Botnet}.
\newblock {\em USENIX Security Symposium}, July 2017.

\bibitem{Aversano:9999bh}
L.~Aversano, M.~L. Bernardi, M.~Cimitile, and R.~Pecori.
\newblock {A systematic review on Deep Learning approaches for IoT security}.
\newblock {\em Computer Science Review}, 40:100389, Jan. 9999.

\bibitem{pack}
Z.~Brakerski, C.~Gentry, and S.~Halevi.
\newblock Packed ciphertexts in lwe-based homomorphic encryption.
\newblock In K.~Kurosawa and G.~Hanaoka, editors, {\em Public-Key Cryptography
  -- PKC 2013}, pages 1--13, Berlin, Heidelberg, 2013. Springer Berlin
  Heidelberg.

\bibitem{etc:motion}
ccrisan.
\newblock {\em MotionEye}, 2020.
\newblock \url{https://github.com/ccrisan/motioneye}.

\bibitem{8629941}
N.~Chaabouni, M.~Mosbah, A.~Zemmari, C.~Sauvignac, and P.~Faruki.
\newblock Network intrusion detection for iot security based on learning
  techniques.
\newblock {\em IEEE Communications Surveys Tutorials}, 21(3):2671--2701, 2019.

\bibitem{secret1}
T.~Chen and S.~Zhong.
\newblock Privacy-preserving backpropagation neural network learning.
\newblock {\em IEEE Transactions on Neural Networks}, 20(10):1554--1564, Oct
  2009.

\bibitem{ckks}
J.~H. Cheon, A.~Kim, M.~Kim, and Y.~Song.
\newblock Homomorphic encryption for arithmetic of approximate numbers.
\newblock In T.~Takagi and T.~Peyrin, editors, {\em Advances in Cryptology --
  ASIACRYPT 2017}, pages 409--437, Cham, 2017. Springer International
  Publishing.

\bibitem{clark2013security}
s.~s. clark.
\newblock {\em the security and privacy implications of energy-proportional
  computing}.
\newblock university of massachusetts amherst, 2013.

\bibitem{clark2013wattsupdoc}
s.~s. clark, b.~ransford, a.~rahmati, s.~guineau, j.~sorber, w.~xu, and k.~fu.
\newblock $\{$wattsupdoc$\}$: power side channels to nonintrusively discover
  untargeted malware on embedded medical devices.
\newblock In {\em 2013 usenix workshop on health information technologies
  (healthtech 13)}, 2013.

\bibitem{web:smpg}
CNet.
\newblock These smart plugs are the secret to a seamless smart home, 2019.

\bibitem{debruin2015powerblade}
s.~debruin, b.~ghena, y.-s. kuo, and p.~dutta.
\newblock powerblade: a low-profile, true-power, plug-through energy meter.
\newblock In {\em proceedings of the 13th acm conference on embedded networked
  sensor systems}, pages 17--29, 2015.

\bibitem{dp}
C.~Dwork, A.~Roth, et~al.
\newblock The algorithmic foundations of differential privacy.
\newblock {\em Foundations and Trends{\textregistered} in Theoretical Computer
  Science}, 9(3--4):211--407, 2014.

\bibitem{gentry2009}
C.~Gentry.
\newblock Fully homomorphic encryption using ideal lattices.
\newblock In {\em Proceedings of the Forty-first Annual ACM Symposium on Theory
  of Computing}, STOC '09, pages 169--178, New York, NY, USA, 2009. ACM.

\bibitem{cryptonets}
R.~Gilad-Bachrach, N.~Dowlin, K.~Laine, K.~Lauter, M.~Naehrig, and J.~Wernsing.
\newblock Cryptonets: Applying neural networks to encrypted data with high
  throughput and accuracy.
\newblock In {\em International Conference on Machine Learning}, pages
  201--210, 2016.

\bibitem{etc:aiy}
google.
\newblock {\em Google AIY Projects}, 2021.
\newblock \url{https://aiyprojects.withgoogle.com}.

\bibitem{Hindle:ug}
A.~Hindle, A.~Wilson, K.~Rasmussen, E.~J. Barlow, J.~C. Campbell, and
  S.~Romansky.
\newblock {Greenminer: A hardware based mining software repositories software
  energy consumption framework}.
\newblock {\em dl.acm.org}, 2014.

\bibitem{etc:mirai}
jgamblin.
\newblock {\em Mirai-Source-Code}, 2017.
\newblock \url{ https://github.com/jgamblin/Mirai-Source-Code}.

\bibitem{jiang2009design}
x.~jiang, s.~dawson haggerty, p.~dutta, and d.~culler.
\newblock design and implementation of a high-fidelity ac metering network.
\newblock In {\em 2009 international conference on information processing in
  sensor networks}, pages 253--264. ieee, 2009.

\bibitem{862209}
P.~Jogalekar and M.~Woodside.
\newblock Evaluating the scalability of distributed systems.
\newblock {\em IEEE Transactions on Parallel and Distributed Systems},
  11(6):589--603, 2000.

\bibitem{Jung-01}
W.~Jung, H.~Zhao, M.~Sun, and G.~Zhou.
\newblock {IoT Botnet Detection via Power Consumption Modeling}.
\newblock In {\em ACM/IEEE CHASE}, 2019.

\bibitem{GAZELLE}
C.~Juvekar, V.~Vaikuntanathan, and A.~Chandrakasan.
\newblock $\{$GAZELLE$\}$: A low latency framework for secure neural network
  inference.
\newblock In {\em 27th $\{$USENIX$\}$ Security Symposium ($\{$USENIX$\}$
  Security 18)}, pages 1651--1669, 2018.

\bibitem{Kim:2008cl}
H.~Kim, J.~Smith, and K.~G. Shin.
\newblock {Detecting energy-greedy anomalies and mobile malware variants.}
\newblock {\em MobiSys}, page 239, 2008.

\bibitem{survvv}
M.~Kuzin, Y.~Shmelev, and V.~Kuskov.
\newblock New trends in the world of iot threats.
\newblock
  \url{https://securelist.com/new-trends-in-the-world-of-iot-threats/87991/},
  2018.

\bibitem{Li:ej}
F.~Li, Y.~Shi, A.~Shinde, and J.~Ye.
\newblock {Enhanced cyber-physical security in internet of things through
  energy auditing}.
\newblock {\em ieeexplore.ieee.org}, 2019.

\bibitem{MiniONN}
J.~Liu, M.~Juuti, Y.~Lu, and N.~Asokan.
\newblock Oblivious neural network predictions via minionn transformations.
\newblock In {\em Proceedings of the 2017 ACM SIGSAC Conference on Computer and
  Communications Security}, pages 619--631, 2017.

\bibitem{Maiti:2019vc}
A.~Maiti and M.~Jadliwala.
\newblock {Light Ears - Information Leakage via Smart Lights.}
\newblock {\em Proceedings of the ACM on Interactive, Mobile, Wearable and
  Ubiquitous Technologies}, 2019.

\bibitem{seal}
microsoft Research.
\newblock {\em Microsoft seal (release 3.2)}, Feb. 2019.
\newblock \url{https:// github.com/Microsoft/SEAL}.

\bibitem{delphi}
P.~Mishra, R.~Lehmkuhl, A.~Srinivasan, W.~Zheng, and R.~A. Popa.
\newblock Delphi: A cryptographic inference service for neural networks.
\newblock In {\em 29th $\{$USENIX$\}$ Security Symposium ($\{$USENIX$\}$
  Security 20)}, pages 2505--2522, 2020.

\bibitem{secret3}
P.~Mohassel and Y.~Zhang.
\newblock Secureml: A system for scalable privacy-preserving machine learning.
\newblock In {\em 2017 IEEE Symposium on Security and Privacy (SP)}, pages
  19--38, May 2017.

\bibitem{msoon}
Monsoon.
\newblock Monsoon power monitor.
\newblock \url{https://www.msoon.com/high-voltage-power-monitor}, 2017.

\bibitem{Myridakis:2021tg}
D.~Myridakis, P.~Myridakis, and A.~Kakarountas.
\newblock {A Power Dissipation Monitoring Circuit for Intrusion Detection and
  Botnet Prevention on IoT Devices.}
\newblock {\em Computation}, 2021.

\bibitem{nazari2017eddie}
a.~nazari, n.~sehatbakhsh, m.~alam, a.~zajic, and m.~prvulovic.
\newblock eddie: em-based detection of deviations in program execution.
\newblock In {\em proceedings of the 44th annual international symposium on
  computer architecture}, pages 333--346, 2017.

\bibitem{Goldreich}
G.~Oded.
\newblock {\em Foundations of Cryptography: Volume 2, Basic Applications}.
\newblock Cambridge University Press, USA, 1st edition, 2009.

\bibitem{or2019dynamic}
o.~or~meir, n.~nissim, y.~elovici, and l.~rokach.
\newblock dynamic malware analysis in the modern era—a state of the art
  survey.
\newblock {\em acm computing surveys (csur)}, 52(5):1--48, 2019.

\bibitem{Panwar:2019uo}
n.~panwar, s.~sharma, s.~mehrotra, l.~krzywiecki, and n.~venkatasubramanian.
\newblock {smart home survey on security and privacy.}
\newblock {\em arxiv.org}, 2019.

\bibitem{iot:phishing}
Radware.
\newblock A game of cat and mouse: Dynamic ip address and cyber attacks, Feb.
  2016.
\newblock \url{https://security.radware.com/
  ddos-threats-attacks/ddos-attack-types/dynamic-ip-address-cyber-attacks}.

\bibitem{sehatbakhsh2018syndrome}
n.~sehatbakhsh, m.~alam, a.~nazari, a.~zajic, and m.~prvulovic.
\newblock syndrome: spectral analysis for anomaly detection on medical iot and
  embedded devices.
\newblock In {\em 2018 ieee international symposium on hardware oriented
  security and trust (host)}, pages 1--8. ieee, 2018.

\bibitem{Shokri}
R.~{Shokri}, M.~{Stronati}, C.~{Song}, and V.~{Shmatikov}.
\newblock Membership inference attacks against machine learning models.
\newblock In {\em 2017 IEEE Symposium on Security and Privacy (SP)}, pages
  3--18, 2017.

\bibitem{Zhang}
F.~Tram{\`e}r, F.~Zhang, A.~Juels, M.~K. Reiter, and T.~Ristenpart.
\newblock Stealing machine learning models via prediction apis.
\newblock In {\em 25th {USENIX} Security Symposium ({USENIX} Security 16)},
  pages 601--618, Austin, TX, Aug. 2016. {USENIX} Association.

\bibitem{pipeline}
Wikipedia.
\newblock Pipeline (computing).
\newblock \url{https://en.wikipedia.org/wiki/Pipeline_(computing)}, 2004.

\bibitem{brute}
Wikipedia.
\newblock Brute-force attack.
\newblock \url{https://en.wikipedia.org/wiki/Brute-force_attack}, 2018.

\bibitem{web:symantec}
Wikipedia.
\newblock 20-year-old flaw found in ubiquiti networking gear running ancient
  php., 2020.

\bibitem{BF}
Wikipedia.
\newblock {\em Brute-Force Attack}, 2020.
\newblock \url{ https://en.wikipedia.org/wiki/ Brute-force\_attack}.

\bibitem{Anonymous:2021gb}
Y.~Xing, H.~Shu, H.~Zhao, D.~Li, and L.~Guo.
\newblock {Survey on Botnet Detection Techniques: Classification, Methods, and
  Evaluation}.
\newblock {\em Hindawi, Mathematical Problems in Engineering}, pages 1--24,
  Apr. 2021.

\bibitem{qi:2018tn}
Q.~Yang, P.~Gasti, K.~S. Balagani, Y.~Li, and G.~Zhou.
\newblock {USB side-channel attack on Tor.}
\newblock {\em Comput. Networks}, 2018.

\bibitem{Yang2017OnIB}
Q.~Yang, P.~Gasti, G.~Zhou, A.~Farajidavar, and K.~S. Balagani.
\newblock On inferring browsing activity on smartphones via usb power analysis
  side-channel.
\newblock {\em IEEE Transactions on Information Forensics and Security},
  12:1056--1066, 2017.

\bibitem{secret2}
J.~Yuan and S.~Yu.
\newblock Privacy preserving back-propagation neural network learning made
  practical with cloud computing.
\newblock {\em IEEE Transactions on Parallel and Distributed Systems},
  25(1):212--221, Jan 2014.

\bibitem{qiao}
Q.~{Zhang}, C.~{Xin}, and H.~{Wu}.
\newblock {SecureTrain}: An approximation-free and computationally efficient
  framework for privacy-preserved neural network training.
\newblock {\em IEEE Transactions on Network Science and Engineering}, (in
  press), URL: https://ieeexplore.ieee.org/document/9271910.

\end{thebibliography}

\end{document}